\documentclass[12pt]{article}
\usepackage{amsmath}
\usepackage{pstricks}
\usepackage{pst-coil}
\usepackage{amsfonts}
\usepackage{epsfig}
\usepackage{citesort}
\usepackage{graphicx, wrapfig}
\def\be{\begin{equation}}
\def\ee{\end{equation}}
\def\bea{\begin{eqnarray}}
\def\eea{\end{eqnarray}}

\def\3tvec#1#2#3{
\left(
\begin{array}{c}
#1  \\
#2  \\
#3  \\
\end{array}
\right)}
\def\lsim{\raise0.3ex\hbox{$\;<$\kern-0.75em\raise-1.1ex\hbox{$\sim\;$}}}
\def\gsim{\raise0.3ex\hbox{$\;>$\kern-0.75em\raise-1.1ex\hbox{$\sim\;$}}}

\begin{document}

\title{
\vglue -0.3cm
\vskip 0.5cm
\Large \bf
Is the tri-bimaximal mixing accidental?}
\author{
{Mohammed Abbas $^{a,b,c}$\thanks{email: \tt
mabbas@ictp.it}~~\,and
\vspace*{0.15cm} ~A. Yu. Smirnov$^{c,d}$\thanks{email:
\tt smirnov@ictp.it}
} \\
{\normalsize\em
$^a$ Ain Shams University,
Faculty of Sciences,} \\
{\normalsize\em Abbassiyah 11566, Cairo, Egypt
 \vspace*{0.15cm}}
\\
{\normalsize\em
$^b$ Center for theoretical physics (CTP), The British University in Egypt, BUE,} \\
{\normalsize\em El-Sherouk City, Cairo, Egypt
\vspace*{0.15cm}}
\\
{\normalsize\em $^{c}$The Abdus Salam International Centre for
Theoretical
Physics,} \\
{\normalsize\em Strada Costiera 11, I-34014 Trieste, Italy
\vspace*{0.15cm}}
\\
{\normalsize\em $^{d}$Institute for Nuclear Research, Russian Academy of
Sciences} \\ {\normalsize\em Moscow, Russia}
}

\input{epsf}
\date{\today}
\maketitle
\thispagestyle{empty}
\vspace{-0.8cm}


\begin{abstract}
\noindent
The Tri-bimaximal (TBM) mixing is not accidental
if structures of the corresponding leptonic mass matrices
follow immediately from certain (residual or broken) flavor symmetry.
We develop a simple formalism which allows one to
analyze effects of deviations of the lepton mixing from
TBM on structure of the neutrino mass matrix
and on underlying flavor symmetry.
We show that possible deviations from the TBM mixing can lead
to strong modifications of the mass matrix
and strong violation of the TBM mass relations.
As a result, the mass matrix may have
an ``anarchical'' structure with random values of elements or
it may have some symmetry which differs from the TBM symmetry.
Interesting examples include  matrices with texture zeros,
matrices with certain ``flavor alignment'' as well as hierarchical
matrices with a two-component structure,
where the dominant and sub-dominant contributions
have different symmetries. This opens up new approaches to
understand the lepton mixing.

\end{abstract}

\section{Introduction}

The lepton mixing determined from the results of neutrino experiments
can be well described by the so called Tri-Bimaximal Mixing (TBM) matrix
\cite{TBM}~\footnote{There is an ambiguity in the form of the mixing matrix
related to the sign of rotation.}:
\bea
U_{TBM}=\left(
\begin{array}{ccc}
            \sqrt\frac{2}{3}  & \frac{1}{\sqrt3} & 0 \\
            -\frac{1}{\sqrt6} & \frac{1}{\sqrt3} & -\frac{1}{\sqrt2} \\
            -\frac{1}{\sqrt6} & \frac{1}{\sqrt3} & \frac{1}{\sqrt2}\\
\end{array}
\right).
\label{tbmmixing}
\eea
In terms of the standard parameterization of lepton mixing matrix,
\bea
U_{PMNS} =  U_{23} (\theta_{23}) \Gamma_\delta U_{13} (\theta_{13})
\Gamma_\delta^* U_{12}(\theta_{12})~,\nonumber
\label{pmns}
\eea
where $\Gamma_\delta \equiv \rm diag (1, 1, e^{i\delta})$,
the TBM matrix  corresponds to maximal
2-3 mixing, zero 1-3 mixing and ``democratic'' 1-2 mixing:
\be
\sin^2 \theta_{23} = \frac{1}{2}, ~~~\sin \theta_{13} = 0, ~~~
\sin^2 \theta_{12} = \frac{1}{3}.
\label{sines}
\ee
The Dirac CP-phase is irrelevant\footnote{In (\ref{pmns})
$U_{ij} \equiv U_{ij} (\theta_{ij})$ is the rotation
in $ij-$ sub-space on the angle $\theta_{ij}$.}.

The result (\ref{tbmmixing}, \ref{sines}) is very suggestive
of certain underlying symmetry and this has triggered enormous activity
in the model-building \cite{Hagedorn:2010th}.
It is assumed that TBM is a consequence of some symmetry
of the neutrino mass matrix in certain (often flavor) basis.
We will refer to this as to the TBM-symmetry.

For the Majorana neutrinos in the flavor basis $(\nu_e, \nu_\mu, \nu_\tau)$, the
mass matrix which leads to the TBM mixing equals
\be
m_{TBM} = U_{TBM}~m_{\nu}^{diag}~U_{TBM}^T,
\label{tbm-m}
\ee
where $m_{\nu}^{diag} \equiv diag(m_1, m_2, m_3)$ is the matrix
of neutrino mass eigenstates. In general, $m_i$ are complex and we can represent them as
\be
m_1 = |m_1|, ~~~ m_2 = |m_2|e^{i2\phi_2}, ~~~  m_3 = |m_3|e^{i2\phi_3}.\nonumber
\ee
Here $\phi_1$ and $\phi_2$ are the Majorana CP-violating phases.
Using (\ref{tbm-m}) and (\ref{tbmmixing}) we find explicitly
\bea
m_{TBM} =
\left(
\begin{array}{ccc}
a    & b   & b \\
...  & \frac{1}{2}(a+b+c) & \frac{1}{2}(a+b-c) \\
...  & ...  & \frac{1}{2}(a+b+c) \\
  \end{array}
\right),
\label{mTBM}
\eea
where the parameters $a,~ b, ~c$ are determined by the neutrino masses as
\be
a = \frac{1}{3}(2m_1 + m_2),
~~~~~~~b=\frac{1}{3}(-m_1 + m_2), ~~~~~~~~c = m_3.
\label{masses}
\ee
Elements of the $\mu\tau-$block of the mass matrix (\ref{mTBM}) equal
\be
a + b + c = \frac{1}{3}m_1 + \frac{2}{3} m_2 + m_3, ~~~
a + b - c = \frac{1}{3}m_1 + \frac{2}{3} m_2 - m_3.\nonumber
\ee

According to (\ref{mTBM}), the elements of matrix,
$||m_{\alpha \beta}||$, $\alpha, \beta = e, \mu, \tau$,
which leads to the TBM mixing,
satisfy  the following three conditions:
\bea
m_{e\mu} & = & m_{e \tau},
\label{first}\\
m_{\mu \mu} & = & m_{\tau \tau},
\label{second}\\
m_{e e} & + & m_{e \mu} = m_{\mu \mu} +  m_{\mu \tau}.
\label{third}
\eea
(The latter is equivalent to $\sum_{\alpha} m_{e\alpha}
= \sum_{\beta} m_{\mu \beta}$.)
Inversely, the mass matrix, which satisfies these relations
leads to the TBM mixing independently of values of neutrino masses.
The form of relation (\ref{third})
changes under the field rephasing: $\nu_e \rightarrow - \nu_e$,
{\it etc.}.
Recall that in the case of bi-maximal mixing
instead of the condition (\ref{third}) we would have
$m_{ee} = m_{\mu \mu} +  m_{\mu \tau}$.

In general fixing any specific set of values of three mixing angles
would imply three relations between the elements of mass matrix.
The point is that in the TBM case these relations are very simple:
they are just equalities of certain elements
and equality of sums of elements of columns,
and therefore have a good chance to follow from certain symmetry.

The TBM symmetry can appear as a residual of the flavor symmetry of
the Lagrangian. (In all the models the underlying
flavor symmetry for TBM is broken.) Indeed, the TBM mass matrix (\ref{mTBM})
is invariant under transformations \cite{lam,grimus}
\be
V_i m_{TBM} V^T_i = m_{TBM},\nonumber
\ee
where
\be
V_1 =
\frac{1}{3} \left(
\begin{array}{lll}
 -1  & ~~ 2 & ~~ 2 \\
 ~ ... & - 1 & ~~ 2 \\
 ~ ... & ... & - 1
\end{array}
\right), ~~~
V_2 =
\left(
\begin{array}{lll}
 1  & 0 & 0 \\
 ... & 0 & 1 \\
 ... & ... & 0
\end{array}
\right).
\label{v1v2}
\ee
At the same time, the  mass matrix of charged leptons
can be diagonal due to symmetry with respect
to transformation $V_3 = diag(1, \omega, \omega^2)$,
where $\omega \equiv e^{i2\pi/3}$. The transformations $V_1, ~v_2, ~v_3$ are generators of the group $S_4$

Some recent developments  have risen
doubts in that the TBM is of fundamental character,
{\it i.e.} follows from certain approximate (broken) symmetry.
The TBM mixing can be  accidental - just
a numerical coincidence of parameters
without underlying symmetry. The arguments follow.

1. Analysis of experimental data shows deviations
from the TBM mixing. According to two recent global analyses
\cite{exp-res}, \cite{Gonzalez},
the  best fit values as well as the $1\sigma$ allowed ranges for
the mixing angles deviate from the TBM values
(see Table \ref{data}).
\begin{table}
\begin{center}
\begin{tabular}{|c|c|c|c|}
  \hline
& Bari group\cite{exp-res} & GM-I \cite{Gonzalez} &
GM-II \cite{Gonzalez}\\
\hline   & & & \\
$\sin \theta_{13}$ & $0.126^{+0.053}_{-0.049}$ &
$0.127^{+0.036}_{-0.055}$&
$0.118^{+0.038}_{-0.048}$ \\ & & &  \\
\hline  &  & & \\
$ \sin^2\theta_{23}$ & $0.466^{+0.073}_{-0.058}$ &$
0.463^{+0.071}_{-0.048} $&
$ 0.463^{+0.071}_{-0.048}$\\ & & & \\
\hline  & & & \\
$\sin^2\theta_{12}$ &$ 0.312^{+0.019}_{-0.018} $&$
0.319^{+0.016}_{-0.016}$ &
$0.321^{+0.016}_{-0.016}$ \\
 &  & & \\
\hline
\end{tabular}
\end{center}
\caption{The best fit values and 1$\sigma$ intervals for the mixing
angles according to global oscillation analysis of different groups.
The analysis GM-I uses the solar neutrino neutrino spectrum
according to the solar model with high metallicity
(GS98) and normal Gallium cross-section, whereas
GM-II is based on the high surface metallicity
(AGSS09) and  modified Gallium cross-section; see
\cite{Gonzalez} for details.
}
\label{data}
\end{table}
Notice, however, that the latest analysis of the atmospheric
neutrino data only
\cite{SKatm} gives the best fit values (and the $90 \%$ CL allowed regions)
as $\sin \theta_{13} = 0.00~ (< 0.2)$ in the case of normal mass
hierarchy (NH) and $\sin \theta_{13} = 0.077 ~(< 0.3)$ for the inverted
mass hierarchy (IH).  So, no significant deviation
of the 1-3 mixing from zero is found,
but the upper bound is in agreement with the global fit results.
For the 2-3 mixing, essentially no deviation from maximal
value is obtained:  $\sin^2 \theta_{23} = 0.50$ (NH)
and $\sin^2 \theta_{23} = 0.53$ (IH). At the same time,
larger deviations from the maximal mixing
are allowed in comparison to the global fit:
$0.407<\sin \theta_{23} < 0.583~(90\%)$ CL.
Comparing the results  of the Table 1 with those in (\ref{sines}),
we find that significant deviations from the TBM
values are allowed.


2. No simple and convincing model for the TBM-mixing has been proposed so far, although the simplest possibilities have been explored
almost systematically. The proposed models have rather complicated structure with large number of assumptions, new elements
(fields) new parameters, {\it ad hoc} quantum number assignments,
and yet additional auxiliary symmetries.
Attempts to realize the proposal ``TBM from symmetry''
can be qualified as the ``symmetry building'' by
introduction and tuning of complicated structure of models.
The mixing does not appear as an {\it immediate}  consequence of symmetry.
On the other hand, if true, this means that there
is rich physics behind observed lepton mixing.

One should add however, that from simple assumption of existence of
discrete symmetry which has irreducible triplet representation
one gets structures which resemble the TBM mixing
but often with the wrong mass spectrum.

3. In most proposed models there is no immediate
relation between the masses and mixing angles and
different physics should be introduced to
explain the mass hierarchies. This is still a matter of opinion and some authors
do not consider lack of the relations
as shortcoming in spite of existence of
the Fritzsch or Gatto-Sartory-Tonin
type relations in the quark sector.

4. The quark sector has small mixing and in the first approximation it can be neglected
so that the quark mixing matrix is diagonal, as a consequence
of certain symmetry. This drastically differs from the lepton mixing and
therefore further complications are required to include
the quark sector into a model.
The Grand unification puts further additional requirements \cite{GUT}.
Of course, it is difficult to expect that quark and lepton mixings are
similar: values of neutrino masses strongly differ from values of quark masses. And furthermore
the neutrino mass may have different nature being of the Majorana type.

5. The quark-lepton complementarity \cite{qlc} with different
underlying physics leads to mixing which is very close to
the TBM mixing.

There are several possible implications of these statements:

\begin{itemize}

\item
The TBM mixing is not accidental in spite of arguments 1) - 5)
and there is certain flavor symmetry behind this mixing.
This symmetry can not be exact symmetry
of the Lagrangian (in the proposed models it is broken spontaneously or
explicitly), and therefore deviations from the TBM mixing at
some level are expected anyway. The deviations can  originate from
(i) renormalization group effects \cite{RGE}, (ii) deviations
from ``correct'' VEV alignment \cite{Barry} \cite{honda}, (iii)
a soft breaking of the $\mu-\tau$ and CP symmetries~\cite{Ge},
(iv)  higher order  corrections of
a flavor symmetry breaking and higher dimensional mass
operators~\cite{Hayakawa}, (v) perturbation of the TBM mass matrix and
contribution from charged lepton sector~\cite{Rodejohann2005}, (vi)
breaking of the mass degeneracy of three heavy (right-handed)
Majorana neutrinos~\cite{zhou}, {\it etc.}.

\item The approximate TBM-mixing is not accidental but
is a manifestation of some other structure or other symmetry
which differs from the flavor symmetries proposed so far for
explanation of TBM. A viable alternatives are the
quark-lepton  complementarity \cite{qlc} and weak complementarity
\cite{wqls}, when the
bi-maximal mixing is obtained as a result of flavor symmetry.

\item The approximate TBM mixing is accidental:
it results from an interplay of different
and to a large extent independent factors or/and contributions.
Some other physics apart from the flavor symmetry is involved.
The mixing results from many step construction and
fixing various parameters by introduction of additional auxiliary
symmetries and structures.

\end{itemize}

The main question we address in the paper is how to disentangle
these possible implications.  Clearly,  the conclusive way to answer
the question is to check predictions of specific models which
explain the TBM mixing. Unfortunately,  most of the proposed
models do not give  new generic or
strict predictions.  Therefore  interpretation of results will be
rather ambiguous. Furthermore, in many cases the underlying physics
is at very high mass scales (GUT or even higher),
so that its direct tests are not possible.

The symmetry, if exists, is realized in terms of mass matrix
and not mixing matrix. Therefore, the step is to explore violation
of the TBM  symmetry of the mass matrix.
If the deviations of the mass matrix from $m_{TBM}$
are large (enhanced), and the symmetry is broken strongly,
the symmetry explanation of the TBM is less plausible.
If in the large region of parameters
(which would correspond to large variety of different structures of
matrix) the mass matrix leads to the approximate TBM mixing,
the TBM looks accidental.

Somewhat similar question  (``is TBM hidden or accidental symmetry?'')
has been discussed in \cite{rodjohann}. In a sense,
the inverse problem has been considered:
small (``soft'') $\sim 20\%$  relative corrections
(perturbations) to the TBM mass matrix elements
have been introduced and consequences of these perturbations
for mixing angles have been studied,
depending on the mass hierarchy and phases. Our approach,
criteria of accidental, and conclusions differ from those obtained
in \cite{rodjohann} (see sect. 4).


The paper is organized as follows.
In sect. 2 we present simple formalism which accounts for the effects
of deviations from the TBM on the structure of neutrino mass matrix.
Using this formalism in sect. 3 we study properties of
the neutrino mass matrices
(in the presence of the deviations) for different mass spectra and
values of the CP-phases. In sect. 4 we consider implications
of the obtained results  for the flavor symmetries. We search  for
some alternative  structures of mass matrix,
and correspondingly, alternative explanation of the observed mixing.
Conclusions are given in sect. 5.  \\

\section{Deviations of the mass matrix from the TBM form}

\subsection{Deviations from the TBM mixing}

Let us define the parameters which characterize the deviation of
mixing angles from the TBM values as
\bea
D_{12}\equiv \frac{1}{3}-s_{12}^2,~~~~~~
D_{23} \equiv \frac{1}{2}-s_{23}^2, ~~~~~~
D_{13}\equiv s_{13},
\label{TBMdeviation}
\eea
where $c_{ij}\equiv \cos\theta_{ij}$ and
$s_{ij}\equiv\sin\theta_{ij}$.
Using results of the Table \ref{data}, we find
the central values and the $1\sigma$ allowed
intervals  of these deviations (see the Table \ref{deviationranges}).
\begin{table}%
\begin{tabular}{|c|c|c|c|c|}
  \hline
Deviation  & Bari group \cite{exp-res} & GM-I \cite{Gonzalez} & GM-II
\cite{Gonzalez} \\
\hline
$\sin \theta_{13}$ & $0.126~(0.077 \div 0.179)$ & $0.127~(0.071 \div 0.163)$
& $ 0.118~(0.069 \div 0.156)$ \\
\hline
 $ D_{23}$ & $0.034~(-0.039 \div 0.092)$ &$ 0.037~(-0.034 \div 0.085)$&$
0.037~(-0.034 \div 0.085) $\\
\hline
$D_{12}$ &$ 0.021~(0.002 \div 0.040) $&$ 0.014~(-0.0016 \div 0.027 )$ & $
0.012~(-0.0036 \div 0.028)$ \\
\hline
\end{tabular}
\caption{Central values and 1$\sigma$ allowed
intervals for the TBM deviation parameters according to the
global analysis of different groups (for more explanation see
caption for the Table 1).
}
\label{deviationranges}
\end{table}
For the 1-2 and 1-3 mixings the relative deviations equal
correspondingly, $3D_{12}$ and $2D_{23}$.
The central values of these deviations and maximal
allowed values at $1\sigma$ level are
$(3 - 6)\%$ and $(6 - 12)\%$ for the 1-2
mixing,
and  $(8 - 10)\%$ and $(18 - 19)\% $ for the 2-3 mixing. Thus,
typical size of the relative deviations is about
$10\%$ for 1-2 mixing and $20\%$ for 2-3 mixing.
The 1-3 mixing
can be compared with values of other mixings:
for central value $s_{13}/s_{12} = 0.23$ and in $1\sigma$ interval:
$s_{13}/s_{12} = 0.33$.
The 1-3 mixing can be  smaller but not much
smaller than other mixings.

Instead of $D_{12}$ and $D_{23}$ we could introduce deviations for
sines:
\bea
d_{12}\equiv \frac{1}{\sqrt{3}} - s_{12},~~~~~~
d_{23} \equiv \frac{1}{\sqrt{2}} - s_{23}.
\label{TBMdeviation}
\eea
In the lowest order there are linear relations between
$d_{ij}$ and $D_{ij}$: $d_{23} = D_{23}/\sqrt{2}$,
$d_{12} = D_{12} \sqrt{3} /2$
in contrast to $s_{13}$, which gives
the deviation from zero. Furthermore, in contrast to $s_{13}$,
the linear deviations $D_{12}, ~D_{23}$ are smaller than quadratic ones.
For the linear deviations we have
$s_{13} \gg d_{12} \sim d_{23}$,
and for the present  best fit values:
\be
s_{13}^2 \sim d_{12} \sim d_{23}.
\label{ordercor}
\ee
It can be a hierarchy of the deviations.

\subsection{Corrections to the neutrino mass matrix}

To account for the effects of deviation from the TBM mixing
on structure of the mass matrix we will perform expansion
of the matrix in powers of the deviation parameters $D_{ij}$.
In the lowest approximation the correction due to $D_{ij}$ equals
\be
U_{TBM} m^{diag} \delta U_{ij}^{(1)T} +   {\rm transponent},
\label{corr-tomat}
\ee
where $\delta U_{ij}^{(1)}$ is the first order
correction to $U_{TBM}$ due to the deviation $D_{ij}$.
Eq.(\ref{corr-tomat}) can be also rewritten in the form
$m_{TBM} U_{TBM} \delta U_j^T +$ transponent.
Because of hierarchy (\ref{ordercor}) we compute
also corrections of the order $s_{13}^2$ which are given by
\be
U_{TBM} m^{diag} \delta U_{13}^{(2)T}
+   U_{13}^{(2)} m^{diag} U_{TBM}^T
+ \delta U_{13}^{(1)} m^{diag} \delta U_{13}^{(1)T}.
\label{second13}
\ee
Here $U_{13}^{(2)}$ is the matrix of second order in $s_{13}^2$.
Using (\ref{corr-tomat})  and (\ref{second13}) we find the mass matrix
in the lowest order approximation as
\bea
m_{\nu}& = &  m_{TBM} +
s_{13}\left(
\begin{array}{ccc}
0 & - \frac{1}{\sqrt{2}} e^{-i\delta} g  &
\frac{1}{\sqrt{2}} e^{-i\delta} g \\
... & \sqrt{2}b e^{i\delta} & 0 \\
... & ... & -\sqrt{2} b e^{i\delta} \\
\end{array}
\right)
\nonumber\\
& + &
\frac{s_{13}^2}{2} \left(
\begin{array}{ccc}
2 e^{-2i \delta} g  & - b & - b \\
... & - g & g \\
... & ...  & - g \\
\end{array}
\right)
\nonumber\\
 & + & D_{23}\left(
\begin{array}{ccc}
0 & b & -b \\
... & a+b-c & 0 \\
... & ... & -(a + b - c) \\
\end{array}
\right)
\nonumber\\
&+&3b ~ D_{12}\left(
\begin{array}{ccc}
-1 & -\frac{1}{4} & -\frac{1}{4} \\
... & \frac{1}{2} & \frac{1}{2} \\
... & ... & \frac{1}{2}\\
\end{array}
\right),
\label{main-eq}
\eea
where
\be
g \equiv c - a e^{2i \delta},
\ee
$a, b, c$ are combinations of the neutrino masses
defined in Eq.(\ref{masses}).
Notice that corrections are proportional to the elements of the
TBM matrix and therefore correlate with the original TBM structure.
It follows from the expression (\ref{main-eq}) immediately that

(i) $s_{13}$ as well as  $D_{23}$ corrections break all three
TBM-conditions (\ref{first} - \ref{third});

(ii) $s_{13}^2$  and $D_{12}$ corrections violate only
the third condition.

Corrections due to non-zero 1-3 mixing depend  of $b$ and combination
$g$ of original parameters $a$ and $c$.

The expression for mass matrix (\ref{main-eq}) can be rewritten
in terms of matrices which explicitly violate
the TBM conditions:
\bea
m_{\nu} & = & m_{TBM} + m_{TBM}^\prime +
\nonumber\\
  & + &   x \left(
\begin{array}{ccc}
0 & 1 & - 1 \\
... & 0 & 0 \\
... & ...  & 0 \\
\end{array}
\right) + y \left(
\begin{array}{ccc}
0 & 0 & 0 \\
... & 1 & 0 \\
... & ... & - 1 \\
\end{array}
\right)
 +  z \left(
\begin{array}{ccc}
1 & 0 & 0 \\
... & 0 & 0 \\
... & ... & 0 \\
\end{array}
\right)~.
\label{main}
\eea
Here $m_{TBM}$ is the original TBM-matrix
(\ref{mTBM}) for a given mass spectrum. The matrix $m_{TBM}^\prime$
has exact TBM-form with the following parameters:
\be
a^\prime  =  \frac{b}{2} \left(\frac{15}{2}D_{12} + s_{13}^2 \right), ~~
b^\prime  =  - \frac{b}{2} \left(\frac{3}{2}D_{12} + s_{13}^2 \right), ~~
c^\prime  = - g s_{13}^2.\nonumber
\ee
Notice that, all the elements of $m_{TBM}^\prime$  are suppressed
in comparison to the zero order matrix $m_{TBM}$ by small
deviations: $D_{12} \sim s_{13}^2 \leq 0.02$.
In Eq.(\ref{main}) $x$ and $y$ are the strengthes of violation of the
first and second TBM-conditions, and $z$ is the correction to
$m_{ee}$:
\bea
x & = & - \frac{s_{13}}{\sqrt{2}} g
e^{-i\delta} + b D_{23},
\label{corr-x}\\
y & = & \sqrt{2} s_{13} b e^{i\delta} + (a + b - c) D_{23} =
\sqrt{2} s_{13} b e^{i\delta} +2 m_{\mu \tau}^{TBM} D_{23},
\label{corr-y}\\
z & = & - \frac{27}{4} b D_{12} + \left[g  e^{-2i\delta}  -
\frac{b}{2}\right] s_{13}^2.
\label{corr-z}
\eea
The corrections to TBM structure have the following properties.
Contributions to $x$ and  $y$ from $s_{13}$ and
$D_{23}$ can sum up, thus enhancing violation of the TBM structure.
All corrections to the elements $m_{e\mu}$ and $m_{e\tau}$
but those of $s_{13}$ are proportional to $b$;
$z$ depends on the smallest deviation $D_{12}$ and second order in
$s_{13}$. In general, parameters $x$ and  $y$ are independent.
If $b \ll a, c$ which, as we will see,
is realized in many situation, then $x \propto s_{13}$, whereas
$y \propto D_{23}$. If $b \sim a, c$,  one can obtain $x \gg y$ or
$x \ll y$ selecting particular value of the phase $\delta$.
In some cases correlation between corrections $x$ and  $y$ and
structure of the original TBM mass matrix appear.

The total correction to the $ee-$ element is
\be
\Delta m_{ee} = a^\prime + z = - 3 b D_{12} +
e^{-2i\delta}g s_{13}^2 .
\label{ee-corr}
\ee
Although $D_{12}$ is small, it enters $\Delta m_{ee}$ with the coefficient
$3$. In other places its effect is small.
Correction to the $\mu \tau-$ element originates from $m_{TBM}^\prime$:
\be
\Delta m_{\mu\tau} =   \frac{3}{2} b D_{12} +
\frac{1}{2} g s_{13}^2 .
\label{mutau-corr}
\ee
It is about 2 times smaller than $\Delta m_{ee}$
and has additional phase difference between the two terms;
$\Delta m_{\mu\tau} = -\frac{1}{2} \Delta m_{ee}$ at $\delta = \pi/2$.
Apart from some special cases this correction is negligible.

Exact expression for the mass matrix  is simplified substantially
if $D_{12} = \delta = 0$:
\bea
m_{\nu} & = &\left(
\begin{array}{ccc}
a  & c_{13} b \sqrt{1 + 2D_{23}}  - \xi_{-} & c_{13}
b \sqrt{1 - 2D_{23}}  + \xi_{+}  \\
 ... & \frac{1}{2}(a+b+c) + y & \frac{1}{2}(a+b-c)\sqrt{1 - 4 D_{23}^2} -
2\sqrt{2} b D_{23} s_{13} \\
 ... & ...  &   \frac{1}{2}(a+b+c) - y \\
  \end{array} \right)
\nonumber\\
 & + & s_{13}^2~(c-a)\left(
\begin{array}{ccc}
1  &  0  &   0 \\
...  &   D_{23}-\frac{1}{2}  &  \frac{1}{2}\sqrt{1 - 4 D_{23}^2} \\
...  &  ... & - D_{23}- \frac{1}{2}    \\
\end{array}\right).
\label{exact-m}
\eea
Here
\bea
\xi_{\pm} & \equiv & \frac{1}{\sqrt{2}}s_{13}c_{13} \sqrt{1 \pm 2
D_{23}}~ (c-a),
\nonumber\\
y & \equiv &  \sqrt{2} b s_{13} \sqrt{1 - 4 D_{23}^2}
+  (a + b - c) D_{23}.
\label{xi-pm}
\eea
The next order corrections, being proportional to $s_{13} D_{23}$,
appear in the off-diagonal
elements:   $m_{\mu \tau}$, $m_{e \mu}$  and $m_{e \tau}$.
From (\ref{xi-pm}) we have
\be
\xi_{\pm} =   \frac{1}{\sqrt{2}}s_{13}c_{13} ~ (c-a)
\pm  \frac{1}{\sqrt{2}} s_{13}c_{13} D_{23}~ (c-a),
\label{next-c}\nonumber
\ee
where the second term gives the same corrections to
$m_{e \mu}$  and $m_{e \tau}$.
In the lowest order we obtain
$$\xi_{+} = \xi_{-} = \xi \equiv  \frac{1}{\sqrt{2}}s_{13}c_{13} (c-a),~~~~
y = D_{23}(a + b - c)  +  \sqrt{2}s_{13} b,$$  so that
\bea
m_{\nu} & = & \left(
\begin{array}{ccc}
a  & b \sqrt{1 + 2D_{23}}  - \xi &
b \sqrt{1 - 2D_{23}}  + \xi  \\
 ... & \frac{1}{2}(a+b+c)+ y & \frac{1}{2}(a+b-c)\\
 ... & ...  &   \frac{1}{2}(a+b+c) - y \\
\end{array} \right).
\label{proposed2}\nonumber
\eea

As follows from the formulas obtained above, modifications
of the matrix depend on structure of the original matrix.
(The latter, in turn, depends strongly on the absolute mass scale,
mass hierarchy and CP-phases. For general dependence
of the mass matrices on CP-phases see \cite{frigerio}).
According to (\ref{main-eq}), corrections are proportional
to the deviations multiplied by different original matrix elements:
$$
\Delta m_{\alpha \beta} =
\sum_{i > j} \sum_{\gamma \delta} f_{\gamma \delta}^{ij} D_{ij} m_{\gamma \delta}, ~~
$$
where $(i, j = 1, 2, 3)$, $(\alpha, \beta, \gamma,  \delta  = e, \mu, \tau)$
and  $f_{\gamma \delta}^{ij}$ are numerical coefficients
which can  contain also the phase factors $e^{i\delta}$ and $e^{-i\delta}$.
Inserting into (\ref{main-eq}) $ a = m_{ee}^0$, $b = m_{e\mu}^0$,
and $c = m_{\mu \mu}^0 - m_{\mu \tau}^0$, we find that the $s_{13}-$corrections
mix the $e-$line and $\mu \tau-$block elements: the corrections
to the $e$-line elements $m_{e \mu}$ and $m_{e \tau}$
are proportional to the elements of $\mu \tau-$block as well as
to $m_{ee}$, whereas the corrections to the $\mu \tau-$block
are proportional to $m_{e \mu}^0$. The $D_{23}-$corections
do not mix elements from different blocks:
$\Delta m_{\mu \mu} = f_{\mu\mu}^{23} D_{23} m_{\mu \tau}$.
The $D_{12}-$ corrections to all elements are proportional
to $m_{e \mu}^0$. $s_{13}^2-$corrections mix the $\mu \tau-$block
elements and $m_{ee}$.
The correction to the subdominant elements can be proportional
to the element of the dominant block and be much larger than the original element.
The elements of the dominant block can get relative corrections
of the order $(20 - 30) \%$ because the corrections
can be enhanced by some additional numerical factors 2 - 3.
In turn, these factors originate from the correction itself
as well as some smallness of the original element (say by factor 1/2 - 1/3).
In the cases when the original flavor matrix has no hierarchy,
the corrections of the order $30 \%$ can lead to ``anarchical''
character of the matrix with random values of elements.

An alternative parameterization  of deviations from
the TBM mass matrix is proposed in \cite{Carone:2009yu} in which the element $m_{ee}$ is unchanged.


\subsection{Basis corrections}

Basis in which the symmetry is introduced may differ
from the flavor basis.
In the symmetry basis, the elements of mass matrix equal
$m_{\alpha \beta}^{(sym)} = m_{\alpha \beta}
+ \Delta m_{\alpha \beta}^b$, where $\Delta m_{\alpha \beta}^b$
is the basis corrections.
Taking in to account mixing in the quark sector
one can assume that the symmetry basis differs from the
flavor basis by the CKM-type rotation. To get some idea
about possible effects we will consider for simplicity
1-2 rotation only, with the
angle $\theta_b$  of the order of Cabibbo angle:
$s_b \equiv \sin \theta_b \sim \sin \theta_C \sim 0.2$.
This rotation gives the following basis corrections:
\bea
\Delta m_{ee}^b & = & -2 s_b c_b m_{e\mu} + s^2_b (m_{\mu\mu} - m_{ee}),
\nonumber\\
\Delta m_{\mu \mu}^b & = & - \Delta m_{ee},
\nonumber\\
\Delta m_{\mu e}^b &  = &  s_b c_b (m_{\mu\mu} - m_{ee}) - 2 s^2_b m_{e\mu},
\nonumber\\
\Delta m_{e\tau}^b &  = &  -  s_b m_{\mu\tau} + (1 - c_b) m_{e\tau} \approx
 -  s_b m_{\mu\tau} +  \frac{s_b^2}{2} m_{e\tau},
\nonumber\\
\Delta m_{\mu \tau}^b & = & s_b m_{\mu\tau} + (1 - c_b) m_{e\tau}
\approx s_b m_{\mu\tau} + \frac{s_b^2}{2} m_{e\tau}.
\eea
Apparently certain correlations between corrections to different elements
exist, especially for some original structures of mass matrix.
For instance, if $m_{\mu e}$ and  $m_{e\tau}$ are small
(as, e.g., in the case of strong normal mass
hierarchy), then  $\Delta m_{ee}^b = \tan \theta_b \Delta m_{\mu e}^b$
corrections to $\Delta m_{e \mu}^b$ and $\Delta m_{e \tau}^b$ are large,
$\Delta m_{e \tau}^b = - \Delta m_{e\mu}^b m_{\mu \tau}/(m_{\mu\mu} -
m_{ee})$,  etc..

Alternatively, the basis corrections can be accounted for by
further deviation of the mixing angles
from their TBM values: $\theta_{ij} \rightarrow \theta_{ij} +
\Delta \theta_{ij}^b$. Therefore, in our consideration
this can be taken  into account by
enlarging possible intervals for $D_{ij}$.
For instance, change of the 1-2 mixing by $\theta_C$ leads to
the interval $\theta_{12} = 20^{\circ} \div 45^{\circ}$.
The upper value corresponds to maximal 1-2 mixing and
the  QLC case. This interval corresponds to
$D_{12} = - 0.17 \div 0.22$.

We will comment on possible additional  changes
of structure of mass matrix due to these corrections.

\subsection{Violation of the TBM conditions}

Violation of the TBM symmetry
of neutrino mass matrix can be characterized by parameters
which describe violation of the  equalities (\ref{first}
-\ref{third}).
For the first two equalities we can introduce
\bea
\Delta_e & \equiv &\frac{ m_{e\mu}-m_{e\tau}}{m_{e\mu}},
\label{vfirst}\\
\Delta_{\mu \tau} & \equiv & \frac{ m_{\mu\mu}-
m_{\tau\tau}}{m_{\tau\tau}}.
\label{vsecond}
\eea
Since the difference $(m_{ee} + m_{e\tau}) -
(m_{\mu\mu} + m_{\mu\tau})$ depends on  $\Delta_e$  and $\Delta_{\mu \tau}$~\footnote{In the
lowest order the difference
equals  $x - y + z - 3bD_{23} \approx m_{e\mu} \Delta_e /2 +
m_{\tau\tau} \Delta_{\mu \tau}/2 + O(D_{12}, s_{13}^2)$.}
we define the third violation parameter in
different way to avoid the strong correlation between the parameters.
The third TBM condition (\ref{third}) can be rewritten
using (\ref{first}) and (\ref{second}) as
$\Sigma_L = \Sigma_R$,
where
\be
\Sigma_L  \equiv   m_{ee}+\frac{m_{e\mu}+m_{e\tau}}{2}, ~~~
\Sigma_R  \equiv  m_{\mu\tau}+\frac{m_{\mu\mu}+m_{\tau\tau}}{2}.\nonumber
\ee
Then the third TBM violation parameter can be introduced as
\be
\Delta_{\Sigma} \equiv  \frac{\Sigma_L - \Sigma_R}{\Sigma_R}.
\label{vthird}
\ee
In $\Delta_{\Sigma}$ effects of large violations of
the 1st and 2nd conditions are excluded.

Specific values of the violation  parameters  correspond to
certain features of the mass matrix.
For instance, $\Delta_e = 1$ corresponds to the texture zero $m_{e\tau} = 0$,
$\Delta_{\mu \tau} \rightarrow \infty$ gives condition for
$m_{\tau\tau} = 0$, {\it etc}..
These values, in
turn, can testify for some new symmetries of the mass matrix.


In what follows we will  express  the TBM- breaking parameters
in terms of $D_{ij}$ and study their dependence on the absolute mass scale,
type of mass spectrum and CP-phases.
We identify situations when the TBM conditions can be strongly violated.
It is convenient to present the diagonal mass matrix
in (\ref{tbm-m}) as
$$
m^{diag} = diag(m_1, m_2, m_3)=m_1~ I~+~ diag(0, m, M),
$$
where $m\equiv m_2-m_1, M \equiv m_3-m_1$ and $I$ is the unit matrix.
For definiteness we will take $s_{13} > 0$.\\

\noindent
{\bf 1. The parameter $\Delta_e$.}
According to (\ref{main}) this parameter can be written as
\be
\Delta_e =
2  \frac{s_{13} + \alpha}{s_{13} -
\tilde{s}_{13} e^{i\tilde{\phi}}} ~,
\label{delta1a}
\ee
where in the first approximation
$\alpha$ and  $\tilde{s}_{13}$ do not depend on $s_{13}$,
and furthermore, $\alpha \propto D_{23}$.
The factor 2 originates from the fact that
$m_{e\mu} - m_{e\tau} = 2 x$, whereas  $m_{e\mu} = x + A$.
The quantity $\tilde{s}_{13} e^{i\tilde{\phi}}$ plays crucial role:
It determines  position of the pole of
$\Delta_e$ which corresponds to texture zero $m_{e\mu} = 0$.
Also it determines values of $s_{13}$ at which some
other special features of the neutrino mass matrix can be  realized.
Indeed, a given value of $\Delta_e$
corresponds to
\be
s_{13} = \frac{\Delta_e \tilde{s}_{13}e^{i\tilde{\phi}}
+ \alpha}{\Delta_e - 2}~.
\label{s13tild}\nonumber
\ee
So, if $\alpha$ is zero or small, which is realized in many cases,
$\tilde{s}_{13}e^{i\tilde{\phi}}$ determines special values of
$\Delta_e$, and correspondingly, special mass relations (see the Table \ref{val-deltae}).
\begin{table}%
\begin{tabular}{|c|c|c|c|c|c|c|}
\hline
$s_{13} e^{-i\tilde{\phi}}/\tilde{s}_{13} $  &  $ -\frac{1}{3}$ &
$\frac{1}{3} $ & $- 1$ & 1 &  $\gg 1 $ \\
\hline

$\Delta_e$ & $\frac{1}{2}$ & $-1$  & 1  & $\infty $ & $\approx 2$ \\

\hline

mass relation  & $2m_{e\tau} = m_{e\mu}$ &
$ m_{e\tau} = 2 m_{e\mu} $ & $m_{e\tau} = 0$ & $m_{e\mu} = 0$ &
$m_{e\tau} = - m_{e\mu}$ \\

\hline
\end{tabular}
\caption{Special values of the violation parameter $\Delta_e$ and the
corresponding relations between elements of the mass matrix.
Here values of the ratio $s_{13}e^{-i\tilde{\phi}}/\tilde{s}_{13} $
are given
for $\alpha = 0$.}
\label{val-deltae}
\end{table}
Which of the possibilities in the Table can be realized
depends on  the upper bound on $s_{13}$ and
value of $\tilde{s}_{13}$, which in turn is given by the mass spectrum and CP-phases.
Realization of possibilities from the left to right in the Table 3 requires
decreasing values of $\tilde{s}_{13}$.\\


In terms of masses and mixing angles
$\Delta_e$ has the following expression
\bea
\Delta_e = \frac{m s_{12}c_{12}(c_{23} - s_{23})-
s_{13} \kappa (c_{23} +s_{23})}{m~s_{12}c_{12}c_{23} - \kappa s_{13}
s_{23}}~,
\label{zerophase1}
\eea
where
\be
\kappa \equiv  M e^{- i\delta}- m s_{12}^2 e^{i\delta} - 2 i m_1 \sin \delta.\nonumber
\ee
Consequently, the pole value and the phase equal
\be
\tilde{s}_{13} \equiv s_{12} c_{12} \cot \theta_{23} \frac{m}{\kappa}
\approx s_{12} c_{12} \frac{m}{\kappa}(1 + 2 D_{23}), ~~~~
\tilde{\phi} \equiv {\rm arg} \left[\frac{m}{\kappa}\right].
\label{tilde-sp}\nonumber
\ee
The expression for $\Delta_e$ can be rewritten approximately as
\be
\Delta_e \approx
2 \frac{s_{13}(1 + D_{23}) - \tilde{s}_{13} D_{23}}{ s_{13} -
\tilde{s}_{13} e^{i\tilde{\phi}}}~.\nonumber
\ee
Then
\be
\alpha \approx (s_{13} - \tilde{s}_{13})D_{23}.
\label{def-alp}\nonumber
\ee

According to (\ref{delta1a}),
$\Delta_e=  1$, which corresponds to $m_{e\tau}=0$, is
realized at
\bea
s_{13}=-(\tilde s_{13}e^{i\tilde\phi}+ 2 \alpha)
=-\frac{\tilde s_{13}(e^{i\tilde\phi}-2D_{23})}{1+2D_{23}}.
\label{deltaeeq1}
\eea
At
\be
s_{13}=\frac{1}{3}(\tilde s_{13}e^{i\tilde\phi}-2\alpha)=\frac{\tilde s_{13}(e^{i\tilde\phi}+2D_{23})}{3+2D_{23}}
\label{deltaeeq-1}\nonumber
\ee
we obtain $m_{e\tau}=2m_{e\mu}$.

The strongest dependence of $\Delta_e$ is on $s_{13}$.
In the case of maximal 2-3 mixing,
$D_{23}=0$,  eq. (\ref{tilde-sp}) gives
\be
\tilde{s}_{13}^0 \equiv  \left|
\frac{s_{12} c_{12} m}{ M e^{- i\delta} -
m s_{12}^2 e^{i\delta}- 2 i m_1 \sin \delta } \right|.
\label{tildes}
\ee
Since the CP phases are unknown, in general,
$\tilde{\phi}$ can take any value. Therefore,
for a given mass hierarchy and $s_{13}$ and  varying
CP-phases, the maximal and minimal values of $\Delta_e$
are realized for  $\tilde{\phi} = 0$ and $ \pi$:
$\Delta_e = \left| 2s_{13}/(s_{13} \pm \tilde{s}_{13}) \right|$.

\begin{figure}[!tbp]
\begin{center}
\includegraphics[scale=1.0]{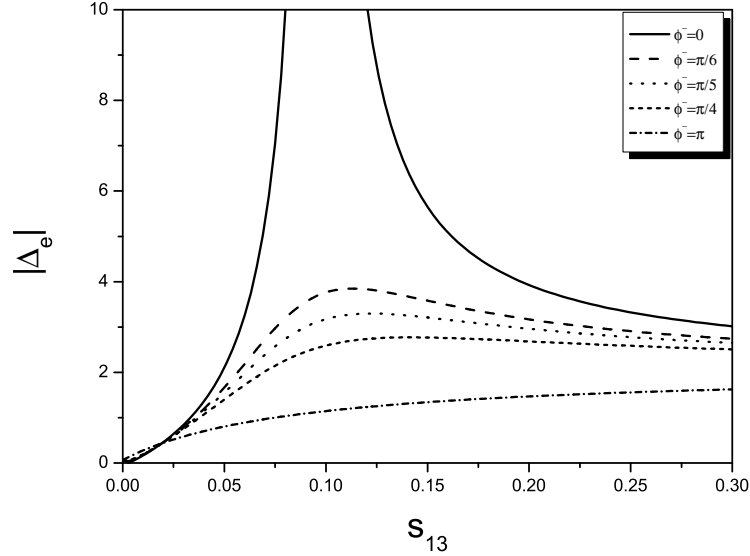}
\end{center}
\caption{$|\Delta_e|$ as a function of $s_{13}$ for different values of
$\tilde{\phi}$. We take
 the best fit values of $\theta_{23}$ and $\theta_{12}$.}
\label{deltaeps}
\end{figure}

If $\tilde{\phi} = 0$, at ${s}_{13} = \tilde{s}_{13}$
, $\Delta_e$ has a singularity.
If $\tilde{\phi} \neq 0$, the function $|\Delta_e|$ has
the peak
\be
|\Delta_e| = \frac{2s_{13}}{\sqrt{(s_{13} -
\tilde{s}_{13} \cos \tilde{\phi})^2
+  (\tilde{s}_{13} \sin \tilde{\phi})^2 } } ~,
\label{delta2a}
\ee
see fig.~\ref{deltaeps}.
The maximum is at $s_{13} \simeq \tilde{s}_{13} \cos \tilde{\phi}$.
For $s_{13} \gg \tilde{s}_{13}$, $\Delta_e$ approaches the asymptotic
value $\Delta_e^{as}= 2$, which corresponds to
the equality $m_{e\mu} = - m_{e\tau}$.

The parameter $\Delta_e$ depends on $m_1$  via $\tilde{s}_{13}$.
As we will see, changing $m_1$ one can increase or decrease $\tilde{s}_{13}$
depending on CP-phases.

According to (\ref{tilde-sp}), a non-zero $D_{23}$ shifts the pole:
$\tilde{s}_{13} =  \tilde{s}_{13}^0 (1 + 2 D_{23})$.
For the present best fit value of $s_{23}$ we obtain
$\tilde{s}_{13} = 1.07 \tilde{s}_{13}^0$,
and for $D_{23}\simeq 0.09$, we have  $\sim 10\%$ change of $\Delta_e$.
The asymptotic value of $\Delta_e$ for large $s_{13}$ becomes
\be
\Delta_e =  1 + \cot \theta_{23} \approx 2  + D_{23}.\nonumber
\ee

In the limit  $s_{13} \rightarrow 0$ we obtain from (\ref{zerophase1})
$\Delta_e = 1 - \tan\theta_{23} \approx 2 D_{23}$.
Then the central and the $1\sigma$
allowed values for $D_{23}$  ($D_{23} = 0.034$ and  $0.09$)
give correspondingly $\Delta_e = (0.07, 0.18)$.

If $D_{23}>0$, the deviation $\Delta_e$ is greater than that in the
case of maximal 2-3 mixing. E.g. in the case of strong mass
hierarchy ($m_1\simeq 0$) and for the best fit values of mixing
angles, we obtain $ \Delta_e \sim 12$ instead of 8.\\

\noindent
{\bf 2. The parameter $\Delta_{\mu \tau}$.}
Similarly to the previous case and according to
eq. (\ref{main}), this violation parameter can be presented
as
\be
\Delta_{\mu\tau} =  -2 \frac{D_{23} + \beta}{D_{23}-\tilde
D_{23}}~,
\label{deltamunu}
\ee
where in the lowest order $\beta$ and the pole value
$\tilde{D}_{23}$ do not depend on  $D_{23}$. In the limit
$\beta \approx 0$, the parameter $\tilde{D}_{23}$ determines
special values of $\Delta_{\mu\tau}$, and consequently,
special relations between the matrix elements (see Table 4).

\begin{figure}[!tbp]
\begin{center}
\includegraphics[scale=1.0]{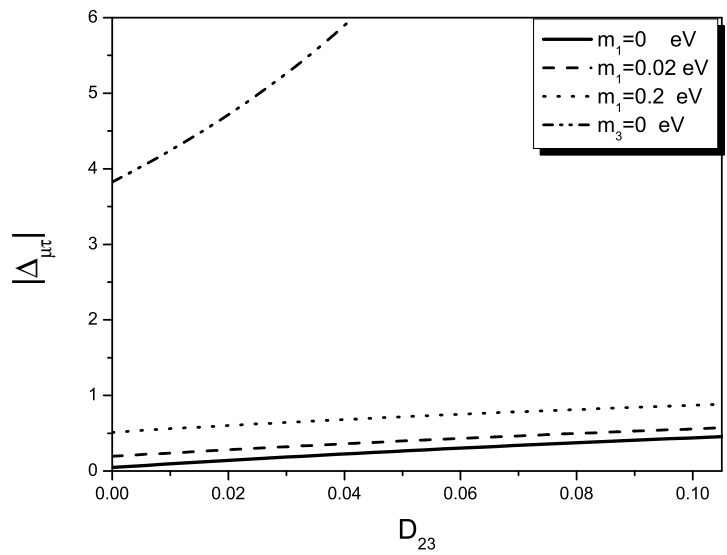}
\end{center}
\caption{Dependence of $|\Delta_{23}|$ on $D_{23}$
for different values of the lightest neutrino mass and $\phi_2=\frac{\pi}{2}$. We take the best fit values of $\theta_{13}$ and $\theta_{12}$. The value $m_3=0$ corresponds to the inverted mass hierarchy.}
  \label{deltamn}
\end{figure}


\begin{table}%
\begin{tabular}{|c|c|c|c|c|c|}
\hline
$D_{23}/\tilde{D}_{23} $  &  $ -\frac{1}{3}$ &
$\frac{1}{3} $ & $- 1$ & 1 &  $\gg 1 $ \\
\hline

$\Delta_{\mu\nu}$ & $- \frac{1}{2}$ & $1$  & $- 1$  & $\infty $ &
$\approx - 2$ \\

\hline

mass relation &
$m_{\tau\tau} = 2 m_{\mu\mu}$ & $ 2m_{\tau\tau} = m_{\mu\mu}$ &
$m_{\mu\mu} = 0$
& $m_{\tau\tau} = 0$ & $m_{\tau\tau} = - m_{\mu\mu}$  \\

\hline
\end{tabular}
\caption{Special values of the violation parameter $\Delta_{\mu\tau}$ and the
corresponding relations between the elements of the mass matrix.
Values of the ratio $D_{23}/\tilde{D}_{23}$
are given
for  $\beta = 0$.}
\label{val-deltamt}
\end{table}


Explicitly, in terms of deviation parameters,
we obtain
\be
\tilde{D}_{23} =
- \frac{1}{2 \kappa_{23}}
\left[ M c_{13}^2 + m c_{12}^2 + 2 m_1 +
2 m^\prime \sqrt{1 - 4D_{23}^2} +
s_{13}^2 \left(m s_{12}^2  e^{2i\delta}
+ m_1 (e^{2i\delta} - 1)\right)\right]\nonumber
\ee
and
\be
\beta = \frac{m^\prime}{\kappa_{23}}\sqrt{1 - 4D_{23}^2} \approx
\frac{m^\prime}{\kappa_{23}},\nonumber
\ee
where
\be
m^\prime  \equiv - m s_{13} s_{12} c_{12} e^{i\delta}\nonumber
\ee
and
\be
\kappa_{23} \equiv M c_{13}^2 - m c_{12}^2 +  m s_{12}^2 s_{13}^2 e^{2i\delta}
+ m_1 s_{13}^2 (e^{2i\delta} - 1).\nonumber
\ee

Neglecting  $s_{13}^2$ terms  we have
in the first approximation
\be
\beta \approx \frac{m^\prime }{\kappa_{23}}=
- \frac{(m_2-m_1)s_{13}s_{12}c_{12}e^{i\delta}}{m_3-m_2c_{12}^2
- m_1s_{12}^2}\nonumber
\ee
and
\be
\tilde{D}_{23} \approx - \frac{m_3 + m_2 c_{12}^2 + m_1 s_{12}^2
-  2 (m_2 - m_1) s_{12}c_{12} s_{13} e^{i\delta}}{2(m_3
- m_2 c_{12}^2 - m_1 s_{12}^2)}~.\nonumber
\ee

For real values of $\tilde{D}_{23}$, this quantity determines
position of the pole of $\Delta_{\mu \tau}$ which corresponds to
$m_{\tau\tau} = 0$. According to (\ref{deltamunu}) the equality
$\Delta_{\mu\tau}= - 1$, $(m_{\mu\mu}=0)$, is
realized at $ D_{23}= -(\tilde D_{23}+ 2 \beta),$
and at
$D_{23}= \frac{1}{3}(\tilde{D}_{23} - 2 \beta)$
we obtain $m_{\mu\mu}
= 2 m_{\tau\tau}$ ($\Delta_{\mu\tau} = 1$). In many situations
$\beta \approx 0$. Non-zero $\beta$ leads to shift of the
special points from values indicated in the  Table 4.

In the lowest order $\Delta_{\mu \tau}$ depends on the 1-3 mixing via
$m^\prime$ only. Neglecting the $s_{13}^2$ corrections, we have
$m_1^\prime  =  m_1$.
The strongest dependence of $\Delta_{\mu \tau}$ is the one on $D_{23}$.
For $s_{13}=0$: we have $m^\prime = 0$, $\beta = 0$ and
\be
\Delta_{\mu \tau}
\approx \left| \frac{ 2D_{23} }
{D_{23} - \tilde{D}_{23}} \right|.
\label{d-mut1}
\ee
In this case
\be
\tilde{D}_{23} =
- \frac{m_3 + m_2 c_{12}^2 + m_1 s_{12}^2 }{2(m_3 -
m_2 c_{12}^2 - m_1 s_{12}^2)}.
\label{d-tilde}
\ee

For maximal  2-3 mixing, $D_{23} = 0$ we obtain from (\ref{deltamunu})
\be
\Delta_{\mu \tau} =
\frac{4m^\prime}{m_1 s_{12}^2 + m_2 c_{12}^2 +  m_3 + 2 m^\prime}.
\label{d-mut13}
\ee
According to (\ref{corr-y}) in the first approximation the
corrections are proportional to the $e\mu-$element
of the original TBM matrix: $\sqrt{2} s_{13} b=\sqrt{2} s_{13}m_{e\mu}^0$.

If  $s_{13}\neq 0$  and $D_{23} \neq 0$ simultaneously,
$\Delta_{\mu \tau}$  can be further enhanced.
The dependence of $\Delta_{\mu \tau}$ on $D_{23}$ is
shown in Fig.\ref{deltamn}.

Notice that the $\mu \tau-$ block
of the mass matrix in all the cases with strong enhancement
of $\Delta_{\mu\tau}$ can be presented as
\bea
m_{\nu} \approx 2 m_0
\left(
\begin{array}{cc}
D_{23} +  \tilde{D}_{23}   &  \frac{1}{2} \sqrt{1 - 4 D_{23}^2} \\
...  &  - D_{23} +  \tilde{D}_{23}
  \end{array}
\right).
\label{mtblocka}
\eea
This shows that when violation of the second condition is strong,
the off-diagonal elements
are much larger (by factor ($2 D_{23})^{-1} >  5$)
than the diagonal elements. In other words,
violation of the TBM condition is large  when
$m_{\mu\mu}$  and $m_{\tau\tau}$  elements are sub-leading.
This means that structure of the whole mass matrix does not change
substantially by these corrections.

The TBM parameters can be introduced in different way:
\be
\Delta_e^\prime  \equiv \frac{m_{e\mu} - m_{e\tau}}{m_{e\mu} +  m_{e\tau} },
\label{vfirst1}
\ee
thus, excluding the linear dependence of the denominator on $s_{13}$. The two parameters are related by
\be
\Delta_e^\prime = \frac{\Delta_e}{2 - \Delta_e}~.\nonumber
\ee
So, that the texture zero $m_{e\mu} = 0$ would correspond to $\Delta_e^\prime = -1$ and the relation $m_{e\mu} = - m_{e\tau}$ is
realized when $\Delta_e^\prime \rightarrow \infty$,  \emph{etc}.. The pole value $\tilde{s}_{13}$ is determined from the
 condition $\Delta_e^\prime (\tilde{s}_{13}) = - 1$.\\

\noindent
{\bf 3.  The parameter $\Delta_{\Sigma}$.}
Using (\ref{main}) we find
\bea
\Sigma_L & = & a + b +
\left(c e^{-2i\delta}- a - \frac{b}{2}\right) s_{13}^2
-  \frac{15}{4} b D_{12},
\nonumber\\
\Sigma_R & = & a + b + 3 b D_{12}~.
\nonumber
\eea
And consequently,
\bea
\Delta_{\Sigma} &\simeq& \frac{s_{13}^2(c
e^{-2i\delta}-a-\frac{b}{2}) - \frac{27}{4} D_{12} b }{a+b+ 3b
D_{12}}\nonumber\\ &=&
\frac{s_{13}^2 \left[m_3
e^{-2i\delta} -\frac{1}{2}(m_1 + m_2)\right] -
\frac{9}{4} D_{12} (m_2 - m_1)
}{\frac{1}{3}m_1 + \frac{2}{3} m_2 + (m_2 - m_1) D_{12}}~.\nonumber
\eea
$\Delta_{\Sigma}$ reflects violation of the
TBM structure by $m_{ee}$ and $m_{\mu \tau}$.
Therefore instead of $\Delta_{\Sigma}$
we can simply use the deviation of  $m_{ee}$
from its TBM value:
\be
\Delta m_{ee} \equiv m_{ee} - m_{ee}^{TBM} =
-  (m_2 - m_1) D_{12} +
\left[ m_3 e^{-2i\delta} - m_1 - (m_2 - m_1)\left(\frac{1}{3} - D_{12}
\right) \right]s_{13}^2.
\label{dmee}\nonumber
\ee
This correction is not affected by the 2-3 mixing.
Contribution of $D_{12}$ is
rather small. Larger effect can be due to $s_{13}^2$.
If $m_1 \approx 0$ the last term can dominate:
$m_{ee} \approx m_3 s_{13}^2$.
Expression (\ref{dmee}) reproduces the one in
(\ref{ee-corr}) when high order terms $ \sim D_{12} s_{13}^2$
are neglected. In the case of strong mass hierarchy
and $s_{13} = 0$ we have $m_{ee} \approx m_2 (1/3 - D_{12})$.


The proposed formalism allows us immediately (and very precisely) to
trace an  impact of deviations from the TBM mixing on structure of
the neutrino mass matrix. Effect of future measurements of the
mixing angles can be seen immediately.

\section{Properties of neutrino mass matrix}

Formulas obtained in the previous section allow us  to ``design''
neutrino mass matrices with certain required  properties which agree
with observations. We reconstruct the neutrino mass matrix in the
cases of TBM mixing and deviations from TBM for different mass
hierarchies and CP-violation phases. Results of numerical
computations are given in the Tables \ref{matr33max} and \ref{matr33}.
The Tables illustrate maximal possible modifications of structures at
certain confidence level. Apparently, any intermediate structure
between the original TBM and  matrices with deviations
presented in the Table \ref{matr33} are possible.
As the best fit values we take $D_{12}=0.012, ~D_{23}=0.037$
and $s_{13}=0.118$ and for $1\sigma$ deviations
we use  $D_{12}=0.028, ~D_{23}=0.085$ and $s_{13}=0.156$.

Modification of the mass matrix (for fixed values of the deviations)
depends on the CP-violating phases.
The Table \ref{matr33} corresponds to $\delta = 0$. For certain cases
this does not correspond to maximal deviation of the
mass matrix from the TBM form. In the Table \ref{matr33max} we show
the mass matrices for $\delta = \pi$ when they lead to stronger
deviations than in Table \ref{matr33}.


Due to hierarchy of the allowed deviations (\ref{ordercor}),
the following combinations of mass matrix elements are
approximately invariant under corrections:
\be
m_{e\mu}  +  m_{e\tau} \approx const, ~~~
m_{\mu \mu}  +  m_{\tau \tau} \approx const.
\ee

The elements $m_{ee}$ and $m_{\mu \tau}$ receive only small corrections.

We will consider several ``benchmark'' spectra determined
by the mass hierarchy/ordering, and CP-parities.
For each case we (i) compute the parameters of mass matrix
and reconstruct the TBM matrix, (ii) find the lowest order
corrections using  (\ref{corr-x} $-$ \ref{corr-z})
and identify conditions at which corrections are maximal,
(iii) compute $\tilde s_{13}, ~\tilde D_{23}$ and the TBM violation parameters,
(iv) discuss properties of the mass matrix with corrections.

\renewcommand{\baselinestretch}{0.5}
\begin{table}
\begin{tabular}{|c|c|c|c|}
  \hline
  Scenario  & Exact TBM & Best fit values & 1 $\sigma$ deviation
\\ \hline & & &\\ NH(0,0) &$\left(
\begin{array}{ccc}
 0.3 & 0.3 & 0.3 \\
 ... & 2.7 & -2.1 \\
 ... & ... & 2.7
\end{array}
\right)$ &$ \left(
\begin{array}{ccc}
 0.35 & -0.06 &  0.7 \\
 ... & ~ 2.6  & -2.1 \\
 ... & ... & ~~ 2.8
\end{array}
\right)$ & \small{$\left(
\begin{array}{ccc}
 0.39 & -0.15 &  0.8 \\
 ... & 2.4 &  -2.0 \\
 ... & ... &  3.0
\end{array}
\right)$} \\\hline & & &\\
  NH$(0,\frac{\pi}{2})$ & $\left(
\begin{array}{ccc}
 0.3 &  0.3 &  0.3 \\
 ... & -2.1 &  2.7 \\
 ... & ... & -2.1
\end{array}
\right)$ & $\left(
\begin{array}{ccc}
 0.2 & 0.7 & -0.16 \\
 ... & -1.8 &  2.7 \\
 ... & ... & -2.4
\end{array}
\right)$& $\left(
\begin{array}{ccc}
 0.14 &  0.8 & -0.34 \\
 ... & -1.5 &  2.6 \\
 ... & ... & -2.6
\end{array}
\right) $\\\hline & & &\\
  PD(0,0)& $\left(
\begin{array}{ccc}
 2.06 & 0.06 &  0.06 \\
 ... & 3.7 & -1.6 \\
 ... & ... & ~ 3.7
\end{array}
\right)$ & $\left(
\begin{array}{ccc}
 2.1 & -0.19 &  0.34 \\
 ... &  3.6 & -1.5 \\
 ... & ... &  3.8
\end{array}
\right)$& $\left(
\begin{array}{ccc}
 2.1 & -0.25 &  0.44 \\
 ... &  3.4 & -1.5 \\
 ... & ... &  3.9
\end{array}
\right)$ \\\hline & & &\\
 PD$(0, \frac{\pi}{2})$ & $\left(
\begin{array}{ccc}
 2.06 &  0.06 &  0.06 \\
 ... & -1.6 &  3.7 \\
 ... & ... & -1.6
\end{array}
\right)$ & $\left(
\begin{array}{ccc}
 1.95 &  0.65 & -0.57 \\
 ... & -1.2 &  3.6 \\
 ... & ... & -1.8
\end{array}
\right)$ & $\left(
\begin{array}{ccc}
 1.9 &  0.79 & -0.81 \\
 ... & -0.86 &  3.5 \\
 ... & ... & -2.1
\end{array}
\right)$ \\\hline & & &\\

  PD$(\frac{\pi}{2}, 0)$ & $\left(
\begin{array}{ccc}
 0.6  & -1.4   & -1.4 \\
 ... &  ~2.2  & -3.0 \\
 ... & ...  & 2.24
\end{array}
\right)$ & $\left(
\begin{array}{ccc}
  0.7 & -1.8  & -0.9 \\
 ...  &  1.7 & -3.0 \\
 ...  & ...  &  2.6
\end{array}
\right)$ & $\left(
\begin{array}{ccc}
  0.8 & -1.9 & -0.7 \\
 ... &  1.3 & -2.9 \\
 ... & ... &  2.9
\end{array}
\right)$ \\\hline & & &\\
  PD$(\frac{\pi}{2}, \frac{\pi}{2})$ & $\left(
\begin{array}{ccc}
 0.6 & -1.4 & -1.4 \\
 ... & -3.0 &  2.2 \\
 ... & ... & -3.0
\end{array}
\right)$ & $\left(
\begin{array}{ccc}
 0.57 & -0.95 & -1.8 \\
 ... & -3.1 &  2.2\\
 ... & ... & -3.0
\end{array}
\right)$ & $\left(
\begin{array}{ccc}
 0.57 & -0.86 & -1.9 \\
 ... & -2.9 &  2.1 \\
 ... & ... & -3.1
\end{array}
\right)$ \\\hline & & &\\
  IH$(0, 0)$ &$ \left(
\begin{array}{ccc}
 4.8 & -0.03 & -0.03 \\
 ... &  2.4 &  2.4 \\
 ... & ... &  2.4
\end{array}
\right)$ & $\left(
\begin{array}{ccc}
 4.8 & 0.36 & -0.44 \\
 ... & 2.6 &  2.4 \\
 ... & ... &  2.3
\end{array}
\right)$ & $\left(
\begin{array}{ccc}
 4.7 & 0.45 & -0.60 \\
 ... & 2.9 &  2.3 \\
 ... & ... &  2.1
\end{array}
\right)$ \\\hline & & &\\
  IH$(\frac{\pi}{2}, 0)$  &
{\small$\left( \begin{array}{ccc}
 1.6 & -3.2 & -3.2 \\
 ... & -0.7 & -0.8 \\
 ... & ... & -0.7
\end{array}
\right)$} & $\left(
\begin{array}{ccc}
 1.7 & -3.2 & -3.2 \\
 ... & -1.4 & -0.8 \\
 ... & ... & -0.2
\end{array}
\right)$& $\left(
\begin{array}{ccc}
 1.9 & -3.2 & -3.1 \\
 ... & -1.7 & -0.8 \\
 ... & ... &  0.05
\end{array}
\right)$
\\\hline & & &\\
D$(0, 0)$& $\left(
\begin{array}{ccc}
 20.0 & 0.006 & ~ 0.006 \\
 ... & 20.3 & -0.3 \\
 ... & ... &  20.3
\end{array}
\right)$ & $\left(
\begin{array}{ccc}
 20.0 & -0.04 & ~ 0.06 \\
 ... & 20.2 & -0.3 \\
 ... & ... &  20.3
\end{array}
\right)$ &$ \left(
\begin{array}{ccc}
 20.0 & -0.05 & ~ 0.07 \\
 ... &  20.2 & -0.3 \\
 ... & ... &  20.3
\end{array}
\right)$ \\\hline & & &\\
  D$(\frac{\pi}{2}, 0)$ & $\left(
\begin{array}{ccc}
 6.6 & -13.3 & -13.3 \\
 ... &  6.9 & -13.6 \\
 ... & ... &  6.9
\end{array}
\right)$ &$ \left(
\begin{array}{ccc}
 7.3 & -14.7 & -11.5 \\
 ... &  3.4 & -13.6 \\
 ... & ... &  9.8
\end{array}
\right)$ &$ \left(
\begin{array}{ccc}
 8.1 & -15.2 & -10.2 \\
 ... &  1.02 & -13.3 \\
 ... & ... &  11.4
\end{array}
\right) $\\\hline & & &\\
  D$(0, \frac{\pi}{2})$ &$ \left(
\begin{array}{ccc}
 20 & ~ 0.006&  0.006 \\
 ... & -0.3 &  20.3 \\
 ... & ... & -0.3
\end{array}
\right)$ &$ \left(
\begin{array}{ccc}
 19.4 & 3.2 & -3.4 \\
 ... & 1.5 &  19.9 \\
 ... & ... & -1.5
\end{array}
\right)$& $\left(
\begin{array}{ccc}
 19.0 & 4.0 & -4.8 \\
 ... & 3.6 &  19.5 \\
 ... & ... & -3.2
\end{array}
\right) $\\\hline & & &\\
  D$(\frac{\pi}{2}, \frac{\pi}{2})$ & $\left(
\begin{array}{ccc}
 6.6 & -13.3 & -13.3 \\
 ... & -13.6 & 6.9 \\
 ... & ... & -13.6
\end{array}
\right)$ & $\left(
\begin{array}{ccc}
 6.8 & -11.4 & -15.0 \\
 ... & -15.4 & 6.7 \\
 ... & ... & -12.0
\end{array}
\right)$ & $\left(
\begin{array}{ccc}
 7.1 & -11.1 & -15.1 \\
 ... & -15.6 & 6.5 \\
 ... & ... & -12.0
\end{array}
\right) $ \\
  \hline
\end{tabular} \caption{Numerical examples of neutrino mass matrices
in the cases of  normal mass hierarchy (NH), partially degenerate spectrum (PD),
inverted hierarchy (IH) and degenerate spectrum (D). The
numbers in brackets of the scenario definition indicate the CP-phases
$(\phi_2, \phi_3)$. We show matrices for the exact TBM (left column),
the best fit values of mixing angles (central column) and mixing
angles allowed at 1$\sigma$ level (right column). We take $\delta=0$ and
the elements of the matrices are in the unit $10^{-2}~$eV. }%
\label{matr33}
\end{table}


\begin{table}
\begin{center}
\begin{tabular}{|c|c|c|c|c|}
  \hline
  Scenario  &  Best fit values &$1\sigma $ deviation
\\ \hline & &\\ NH$(0,0)$ &$\left(
\begin{array}{lll}
 0.35 & 0.67 & -0.11 \\
 ... & 2.5 & -2.1 \\
 ... & ... & ~~2.9
\end{array}
\right)$& $\left(
\begin{array}{lll}
 0.38 & 0.77 & -0.28 \\
 ... & 2.3 & -2.0 \\
 ...& ... & ~~3.1
\end{array}
\right) $\\\hline & &\\
  PD$(0,0)$&  $\left(
\begin{array}{lll}
 2.1 & 0.32 & -0.21 \\
 ... & 3.5 & -1.55 \\
 ... & ... & ~~3.8
\end{array}
\right)$ & $\left(
\begin{array}{lll}
 2.1 & 0.38 & -0.32 \\
 ... & 3.4 & -1.5 \\
 ... & ... & ~~3.9
\end{array}
\right)$\\\hline & & \\
  PD$(\frac{\pi}{2}, \frac{\pi}{2})$ & $ \left(
\begin{array}{lll}
 0.57 & -1.9 & -0.81 \\
 ... & -2.6 & ~~2.1 \\
... & ... & -3.4
\end{array}
\right)$ & $\left(
\begin{array}{lll}
 0.57 & -2.05 & -0.51 \\
 ... & -2.4 & ~~2.0 \\
 ... & ... & -3.7
\end{array}
\right)$\\\hline& &\\
  IH$(0, 0)$ & $\left(
\begin{array}{lll}
 4.8 & -0.41 & 0.39 \\
 ... & ~~2.6 & 2.4 \\
 ... & ... & 2.3
\end{array}
\right)$ &$\left(
\begin{array}{lll}
 4.7 & -0.51 & 0.55 \\
 ... & ~~2.9 & 2.3 \\
... & ... & 2.1
\end{array}
\right)$\\\hline & &\\
  IH$(\frac{\pi}{2}, 0)$  &$
 \left(
\begin{array}{lll}
 1.75 & -3.44 & -2.91\\
 ... & -0.36 & -0.9 \\
 ... & ... & -1.3
\end{array}
\right)$&$\left(
\begin{array}{lll}
 1.9 & -3.6 & -2.6 \\
... & -0.37 & -1.05 \\
 ... & ... & -1.4
\end{array}
\right)$
\\\hline &&\\
  D$(\frac{\pi}{2}, \frac{\pi}{2})$ &  $\left(
\begin{array}{lll}
 6.7 & -15.8 & -10.2 \\
... & -11.0 & ~~6.3 \\
 ... & ... & -16.3
\end{array}
\right)$& $\left(
\begin{array}{lll}
 7.1 & -16.7 & -8.4 \\
 ... & -10.0 & ~~5.5 \\
 ... & ... & -17.7
\end{array}
\right)$\\
  \hline
\end{tabular}
\end{center}
\caption{The same as in Table \ref{matr33} for the Dirac phase $\delta=\pi$.}
\label{matr33max}
\end{table}


\subsection{Normal mass hierarchy }

In the case of strong normal mass hierarchy, we take $m_1 \approx 0$; see lines $NH(0,0)$ and $NH(0,\frac{\pi}{2})$ in the Table \ref{matr33}.\\

1). The parameters of mass matrix\\
  $$a = b \approx \frac{m_2}{3} \approx \frac{\sqrt{\Delta m^2_{21}}}{3},~~
~~~a, b \ll c  \approx m_3 \approx \sqrt{\Delta m^2_{31}}$$ give the TBM matrix
\bea
m_{\nu} & \approx & \left(
\begin{array}{ccc}
\frac{m_2}{3}  &  \frac{m_2}{3}  & \frac{m_2}{3}  \\
 ... & \frac{m_3}{2} + \frac{m_2}{3} & - \frac{m_3}{2} + \frac{m_2}{3}\\
 ... & ...  &   \frac{m_3}{2} + \frac{m_2}{3}\\
\end{array} \right). \label{mtbm-h} \nonumber
\eea

2). The lowest order corrections equal\\
\bea%
x &\approx& - \frac{1}{\sqrt{2}} s_{13} m_3 e^{-i\delta},
~~~ y \approx  - D_{23} m_3 + \frac{\sqrt{2}}{3} s_{13} m_2
e^{i\delta}, \label{corr-nh}\nonumber\\
\Delta m_{ee} &\approx& - m_2 D_{12} + s_{13}^2 m_3
e^{-2 i \delta},~~~~~~~
\Delta m_{\mu \tau}
\approx \frac{1}{2}(m_2D_{12}+m_3s_{13}^2).
\eea%
Notice that in $y$ the two contributions can be of the same size and
enhance each other. The same is in $\Delta m_{ee}$ and $\Delta
m_{\mu \tau}$. For
$D_{23} > 0$, maximal deviations are achieved if $\phi_3 = \pi/2$
and $\delta  = 0$ or $\phi_3 = 0$ and $\delta  = \pi$. For the best
fit values of mixing angles, the maximal deviations equal (in the
units $10^{-2}$ eV) $|\Delta m_{ee}| \sim 0.10$, $|x| \sim 0.45$, $|y| \sim
0.25$, and the correction to the sub-leading elements are bigger
than the original TBM elements. At $1\sigma$ level the corrections
become $|\Delta m_{ee}| \approx 0.15$, $|x| \approx 0.65$, $|y| \approx
0.5$ and structure of
the mass matrix can substantially deviate from the TBM form.\\

3). The parameters of violation of the TBM conditions:
At $D_{23} = 0$, we have
\be
\tilde{s}_{13}^0 \approx s_{12} c_{12} \sqrt{\frac{\Delta
m^2_{21}}{\Delta m_{31}^2}}
\simeq 0.09, \label{tildha}\nonumber
\ee%
and $\tilde{\phi} \approx 2\phi_2 - 2\phi_3  + \delta$.
Notice that $\tilde{s}_{13}^0$  in (\ref{tildha}) is slightly smaller than the
present best fit value of $s_{13}$ and at $1\sigma$ level
$s_{13}/\tilde{s}_{13} \leq 2$. Therefore all the possibilities
indicated in the Table \ref{val-deltae} can be realized.
For the best fit value of 1-3 mixing:
$\Delta_e = 11.6$. For the $1\sigma$ upper bounds on 1-3
mixing, the parameter equals $\Delta_e = 6.4$.
Thus, the first TBM-relation
in (\ref{first}) can be broken very strongly. Such a strong
influence (even for small $s_{13}$) originates from the fact that
$s_{13}$ mixes the large and small mass scales in the mass matrix,
and therefore the corrections to the sub-leading elements
($m_{e\mu}$, $m_{e\tau}$) are proportional to  the large mass: $\sim
s_{13} \sqrt{\Delta m_{31}^2}$.

From (\ref{d-tilde}) we have
\be%
\tilde{D}_{23} =  - 1/2, ~~~ \beta \approx s_{13} s_{12}c_{12}
\frac{m}{M} \ll \tilde{D}_{23},\nonumber
\ee%
and therefore
\be%
\Delta_{\mu \tau} \approx \frac{4 D_{23} }{1 + 2 D_{23}}
\approx 4 D_{23}.\nonumber%
\ee 
Since $D_{23}/\tilde{D}_{23} < 0.2$ ($1\sigma$), no texture zeros or
special relations indicated in the Table \ref{val-deltamt} can be
obtained. Effect
of 1-3 mixing is very small,  since the element $b$ is small. 
According to (\ref{d-mut13}): $\Delta_{\mu \tau} \approx  2 s_{13}
\sin 2 \theta_{12} \sqrt{\Delta m_{21}^2/\Delta m_{31}^2} \sim
s_{13}/3$.

The examples of the Table \ref{matr33} correspond to $\delta =
0$. For $\delta = \pi$, according to eq. (\ref{corr-nh}), the  values
of $m_{e\mu}$ and $m_{e\tau}$ permute, see table \ref{matr33}. Also in this case $m_{ee}$ is
suppressed. Signs of corrections to the $\mu\tau-$ block and $e-$
line elements can be independently changed varying  $\phi_3$ and
$\delta$. Correction to $m_{ee}$ is then fixed.\\

4). Properties of the mass matrix:
\begin{itemize}
\item The allowed corrections to the sub-leading $e\mu-$ and $e\tau-$ elements
dominate the original TBM values: $x \gg b$; changes of elements
of the $\mu\tau-$block can be of the order 1; $m_{ee}$ can be
suppressed by the corrections of the order $s_{13}^2$.


\item
Texture zeros appear: $m_{e\tau} = 0$ or $m_{e\mu} = 0$
at $s_{13}$ determined by $\tilde{s}_{13}$.

\item
Special relations $m_{e\mu} = r m_{e\mu}$, with $r = 1/2, 2$
can be obtained.

\item
The equality $m_{ee} = - m_{e\mu}$ can be approximately realized.

\item Sharp difference of the elements of the $\mu \tau-$ block and
the $e-$line disappears. So, one may have a smooth decrease of
values of the elements from $m_{\tau\tau}$ to $m_{ee}$ with
additional smallness of $m_{e\tau}$. This structure resembles the
structure of the quark mass matrices with, however, much larger
expansion parameter $\lambda \sim 0.5 - 0.8$.

\item Maximal deviation of $m_\nu$ from $m_{TBM}$ corresponds to $m_2
> 0$, $m_3 > 0$ and $\delta = \pi$, which leads to strong increase
of $m_{e\mu}$ and decrease of $m_{\mu \mu}$. In this case correction
to $m_{ee}$ is positive. The $ee-$ element is suppressed, if
$m_2 > 0$, $m_3 < 0$ and $\delta = 0$. In this case the mass matrix
has the following form \be m_\nu = \left( \begin{array}{lll}
 0.4 & 0.8 & ~0.2 \\
 ... & 2.3 & -2.0 \\
 ... & ... &  ~3.0
\end{array}
\right) 10^{-2} ~{\rm eV}.\nonumber
\ee

\end{itemize}
The basis corrections can further smear difference of the
$e$-line and $\mu\tau-$ block elements. Varying $S_b$ in the interval $-0.2 \div 0.2$ one finds $\Delta
m_{ee}^b = (-0.24 \div 0.4)$, $\Delta m_{e\mu}^b = (0.32 \div -
0.44)~$, $\Delta m_{e\tau}^b = (0.4 \div - 0.4)$ in the units $~10^{-2}$ eV.

Total correction to $m_{ee}$ can be as large as $0.002$ eV which is
still smaller than the original $a = 0.003$ eV. However, $m_{ee}= 0$
can be realized with increase of $m_1$. This can be achieved
if $$m_1=-\frac{m_2}{2}+\frac{9}{4c_{13}^2}m_2D_{12} +
\frac{3}{2}m_3\tan^2 \theta_{13}.$$
Numerically, $m_{ee}= 0$ if
$m_1\approx 5.2~. ~10^{-3}$ eV for TBM case, $m_1\approx 3.3~. ~10^{-3}$ eV and
$m_1\approx 6~. ~10^{-3}$ eV for the best fit values of mixing angles with
$\delta=0, ~\frac{\pi}{2}$ respectively.


\subsection{Partially degenerate spectrum}

Suppose $|m_1| \approx |m_2| \approx \bar{m} < |m_3|$. Numerically
this  corresponds to $\bar{m} \sim (2 - 3)\cdot 10^{-2}$ eV and $m_3
= (5.5 - 6.0) \cdot 10^{-2}$ eV. The phase $\phi_2$ becomes
important.\\

\noindent
A. The case $\phi_2 = 0$, lines $PD(0, 0)$ and
$PD(0, \frac{\pi}{2})$ in the Table \ref{matr33}.\\

1). The parameters of the mass matrix
\be
m_1 \approx m_2 \approx \bar{m} >
0, ~~a \approx \bar{m}, ~~
b = \epsilon \equiv
\frac{\Delta m_{21}^2}{6 \bar{m}} \approx 7 \cdot 10^{-4}~ {\rm eV},
\nonumber
\label{eps-s}
\ee%
give the TBM mass matrix
\bea%
m_{\nu} & \approx
& \left( \begin{array}{ccc}
\bar{m}  &  \epsilon  & \epsilon  \\
 ... & \frac{1}{2}(m_3 + \bar{m}) & - \frac{1}{2}(m_3 - \bar{m})   \\
 ... & ...  &   \frac{1}{2}(m_3 + \bar{m}) \\
\end{array} \right).\nonumber
\label{mtbm-ph}
\eea%
The main feature of this matrix is strong (factor of 30) suppression
of the $m_{e\mu}$ and $m_{e\tau}$ elements in comparison with the
other elements which are of the same order. Now $\bar{m} \sim m_3$,
and consequently, strong difference of $m_{\mu \mu}$ and
$m_{\mu\tau}$ can appear.\\

2). The lowest order corrections equal\\
 \be
x = - \frac{1}{\sqrt{2}} s_{13}
\left( m_3 e^{-i\delta} - \bar{m} e^{i\delta} \right),~~~
y =   D_{23} (\bar{m} - m_3), ~~~
\Delta m_{ee} =
s_{13}^2 \left( m_3 e^{-2 i \delta} - \bar{m}  \right).\nonumber
\ee
In the case of $PD(0,0)$ the corrections are not large: $m_3$ and
$\bar{m}$ terms partially cancel each other in $x$ and $y$. Although
$x \gg \epsilon$, the elements $m_{e\mu}$ and $m_{e\tau}$ are small
and structure of the matrix with the dominant $\mu\tau-$ block
does not change.

The situation is different for $\phi_3 = \pi/2$, see line
$PD(0,\frac{\pi}{2})$ of the Table \ref{matr33}. Corrections are
maximal if $m_3 < 0$  and  $\delta = 0$: $$x = - \frac{1}{\sqrt{2}}
s_{13} ( |m_3|  + \bar{m}), ~~~y = D_{23} (\bar{m} + |m_3|), ~~~\Delta
m_{ee} = - s_{13}^2 (|m_3| + \bar{m}).$$ For $\delta = \pi$ the
correction $x$ changes the sign and
values of $m_{e\mu}$ and $m_{e\tau}$ interchange.\\

3). Violation of the TBM conditions:
  According to
(\ref{tilde-sp})
\be%
\tilde{s}_{13} = s_{12}c_{12}  \frac{\Delta
m_{21}^2}{2 \bar{m} \kappa} \sim 10^{-2},\nonumber
\ee%
so that for maximal allowed $s_{13}^{max}$, we have $s_{13}^{max}/\tilde{s}_{13} \sim
20$ and therefore all special mass relations of the Table \ref{val-deltae} can
be satisfied. $\alpha = (s_{13} - \tilde{s}_{13})D_{23},$ and since
$s_{13}^{max} \gg \tilde{s}_{13}$, the corrections of the order
${s}_{13} D_{23}$ become important.

From (\ref{d-tilde}) we find $\tilde{D}_{23} = - (m_3 + \bar
m)/2(m_3 - \bar m) \sim 3/2$ which is larger than in the case of
strong mass hierarchy, and correspondingly,
effect of violation of the 2nd condition is weaker.
In this case $\beta \approx 0$.\\

4). Properties of mass matrix $PD\left(0, \frac{\pi}{2}\right)$:\\
\begin{itemize}
\item Corrections to the sub-leading elements are large: about order
of magnitude larger than the TBM values. Therefore the sub-leading
mass matrix can be modified completely.

\item
At $1 \sigma$ level  the mass matrix has all elements
of the same order (within factor of 3). This can be considered as a
realization of the anarchical structure.

\item
Equality $m_{e\mu} \approx - m_{e\tau}$ can be achieved.
Exact zero of one of these elements
is realized for very small $s_{13}$. Equalities
$m_{e\mu} = - m_{\mu \mu}$ or $m_{ee} = - m_{\mu \mu}$
can be obtained.

\end{itemize}
Basis corrections can further ``equilibrate'' elements. For $s_b=0.2$, they equal $\Delta m_{ee}^b=0.04 $,
$\Delta m_{\mu e}^b=0.31$,
$\Delta m_{e \tau}^b = 0.32 $
and $\Delta m_{\mu \tau}^b=-0.31 $ in the units $10^{-2}$ eV. They are of the order of the TBM violation corrections for the $ee-$ and $e\tau-$ elements.\\

\noindent
B. $\phi_2 = \pi/2$; see line $PD(\frac{\pi}{2}, 0)$
and $PD(\frac{\pi}{2}, \frac{\pi}{2})$.\\

1). The parameters of the mass matrix
$$m_1 \approx \bar{m},~~~~~
m_2 \approx - \bar{m}, ~~~~~a \approx \frac{\bar{m}}{3}, ~~~~~~b
\approx - \frac{2\bar{m}}{3}$$  lead to the TBM mass matrix
\bea m_{\nu} &
\approx & \left( \begin{array}{ccc}
\frac{1}{3} \bar{m}  &  - \frac{2}{3} \bar{m}  & - \frac{2}{3} \bar{m}\\
 ... & \frac{m_3}{2} - \frac{\bar{m}}{6} &
- \frac{m_3}{2} - \frac{\bar{m}}{6}    \\
 ... & ...  &  \frac{m_3}{2} - \frac{\bar{m}}{6} \\
\end{array} \right).
\label{mtbm-ph2}
\eea%
 All the elements are
of the same order, so that corrections do not change the structure
strongly. The $ee-$ element  is the smallest one.\\

2). The lowest order corrections equal
\bea%
x  &=&  - \frac{1}{\sqrt{2}} s_{13} m_3 e^{-i\delta}
 - \frac{2}{3}\bar{m} D_{23},~~~~~
y  =  - \frac{2\sqrt{2}}{3} s_{13} \bar{m} e^{i\delta} - D_{23}
\left(\frac{1}{3}\bar{m} + m_3 \right),\nonumber\\
\Delta m_{ee}&=&s_{13}^2 m_3  + 2\bar{m} D_{12},~~~~~~~~~~~
\Delta m_{\mu\tau} = \frac{1}{2} \Delta m_{ee}-\bar{m}D_{12}.\nonumber
\eea%
For $D_{23} > 0 $ the largest deviations appear when  $\delta  = 0$
and $m_3 > 0$ ($\phi_3 = 0$) or $\delta  = \pi$ and $\phi_3 <
\pi/2$. For the bf-values of mixing parameters (and $1 \sigma$) we
have $x \sim - 0.5 ~(- 0.8)$, $y \sim -0.3~ (- 0.7)$, $\Delta m_{ee}
\sim 0.15~ (0.3)$ (in the units $10^{-2}$ eV). Corrections to the
$e\mu-$ and $e\tau-$ elements are of the order 1; corrections to
other elements are up to $20 - 30 \%$.\\

3). The parameters of violation of the TBM conditions:
   The poles of $\Delta_e$ and $\Delta_{\mu\tau}$ are at
 $$\tilde{s}_{13} =\frac{-\bar{m}\sin 2\theta_{12}(1+2D_{23})}{M
e^{-i\delta}+2\bar{m}s_{12}^2e^{i\delta}-2im_1\sin \delta},$$
$$\tilde D_{23}= -\frac{m_3- \bar{m}(1-s_{13}\tan
2\theta_{12}e^{i\delta})}{2(m_3+\bar m)} \gg
D_{23}^{max},$$ where $D_{23}^{max}$ is maximal
allowed value of $D_{23}$ at $1\sigma$, so no texture zeros
are realized in the $\mu\tau- $ block.\\

\noindent
4). Properties of mass matrix:
\begin{itemize}
\item
texture zero $m_{e\tau} = 0$, is realized at
$s_{13}\approx(0.13, 0.19, 0.34)$ for $m_1=(0.005, 0.01, 0.02)$eV;

\item
in the case of $PD\left(\frac{\pi}{2}, 0\right)$
the structure is possible with all
elements being of the same order
 and $m_{ee}$ and $m_{e\tau}$ being the smallest ones.

\end{itemize}
The basis corrections for $s_b=0.2$ equal $\Delta m_{ee}^b=0.6 $,
$\Delta m_{\mu e}^b=0.43$,
$\Delta m_{e \tau}^b = 0.58 $
and $\Delta m_{\mu \tau}^b=-0.63 $ in the units $10^{-2}$ eV. They are of the order of the TBM violation corrections for the $e\mu-$ and $e\tau-$ elements and large for the $ee-$ and $\mu\tau-$ elements.\\

\subsection{Inverted mass hierarchy}

If  $m_3 \approx 0$, we obtain  $|m_1| \approx |m_2| \approx \bar{m}$,
$\bar{m} \sim \sqrt{\Delta m_{31}^2} \approx  5 \cdot 10^{-2}$ eV.
Structure of the mass matrix is similar
to that in the partially degenerate case.
Similarly, the results strongly depend on the phase $\phi_2$.\\

\noindent
A. $\phi_2 = 0$, line $IH(0,0)$.\\

1). Parameters of the mass matrix\\
$$m_1 \approx m_2 \approx \bar{m} > 0,~~~
 a \approx \bar{m},~~~
b = \epsilon  \sim 2.7 \cdot 10^{-4}~ {\rm eV},
$$

give the TBM mass matrix
\bea
m_{\nu} & \approx & \left(
\begin{array}{ccc}
\bar{m}  &  \epsilon  & \epsilon  \\
 ... & \frac{1}{2}\bar{m} & \frac{1}{2}\bar{m}   \\
 ... & ...  &   \frac{1}{2} \bar{m} \\
\end{array} \right).\nonumber
\label{mtbm-inv1}
\eea
The elements $m_{e\mu}$ and $m_{e\tau}$ are suppressed
by 2 orders of magnitude in comparison to the other elements.\\

2). The lowest order correction:
\be
x =  \frac{1}{\sqrt{2}} s_{13} \bar{m} e^{i\delta},~~~
y =   D_{23} \bar{m}, ~~~
\Delta m_{ee} = - s_{13}^2 \bar{m},~~~\Delta m_{\mu\tau} = \frac{1}{2}\Delta m_{ee}.\nonumber
\ee
 Corrections strongly correlate with the TBM structure:
they suppress the  $ee-$ and $\tau\tau-$ elements, and enhance
the $\mu\mu-$ element (if $D_{23} > 0$). Corrections to
$m_{e\mu}$ and $m_{e\tau}$ dominate, so that
$m_{e\mu} \approx - m_{e\tau}$. The matrix with corrections can be written as
\bea%
m_{\nu} \approx
\bar{m} \left(
\begin{array}{ccc}
1 - s_{13}^2 &  \frac{1}{\sqrt{2}}s_{13} e^{i\delta} &
-\frac{1}{\sqrt{2}} s_{13} e^{i\delta} \\
... & \frac{1}{2}  + D_{23} & \frac{1}{2} \\
... & ... & \frac{1}{2} - D_{23} \\
\end{array}
\right).
\label{mtot5}\nonumber
\eea%
Corrections are small to $m_{ee}$ and at the  bf-values ($1\sigma$ level) of the deviation
they equal approximately
$10\%$ $(20\%)$ for elements of the $\mu \tau-$ block.\\

3). The parameters of violation of the TBM conditions:
  If $D_{23} = 0$,  we obtain from (\ref{tildes}) very small pole value
\be
\tilde{s}_{13}
\approx
s_{12} c_{12} ~\frac{\Delta m_{21}^2}{2 \Delta m_{31}^2}  \approx 0.008.
\nonumber
\ee
Consequently, for the central values of the 1-3 mixing we  have
nearly maximal TBM violation, $\Delta_e \approx 2$.
All the relations in the Table \ref{val-deltae} can be satisfied.

For  $s_{13} = 0$ we have
$
\tilde{D}_{23} \approx 0.5 ,
$
and as can be immediately seen from Eq. (\ref{mtot5}),
\be
\Delta_{\mu \tau} =
 4 D_{23} \frac{1}{1 - 2 D_{23}}~ \approx 4 D_{23},
\label{d-mutau5a}\nonumber
\ee
independently of the phase $\phi_2$.\\
If $D_{23} = 0$ we obtain from (\ref{d-mut13})
$
\Delta_{\mu \tau}
\approx
s_{13} \sin 2 \theta_{12}\frac{\Delta m_{21}^2}{\Delta m_{31}^2}
$
which is strongly suppressed.\\

4). Properties of the mass matrix, (line $IH(0,0)$):\\
\begin{itemize}

\item

Structure of the dominant block of the mass matrix does not change substantially
in comparison to the TBM form.

\item
No texture zeros can be obtained in the $\mu \tau-$ block.

\item
Matrix has no special
structure  apart from some trend of increase of elements from
$m_{\tau \tau}$ to $m_{ee}$, with $m_{ee}$ being the largest
one.

\item
Equality $m_{e\mu}\approx -m_{e\tau}$ can be achieved.
Texture zeros $m_{e\mu}$ or $m_{e\tau}$ are possible for very small
values of $s_{13}$.

\end{itemize}

The basis corrections for $s_b=0.2$ equal $\Delta m_{ee}^b=-0.08 $,
$\Delta m_{\mu e}^b=-0.47$,
$\Delta m_{e \tau}^b = -0.48 $
and $\Delta m_{\mu \tau}^b=0.48 $ in the units $10^{-2}$ eV. They are of the order of the TBM corrections for $e\mu-$ and $e\tau-$ elements.\\

\noindent
B. $\phi_2 = \pi/2$, line $IH(\frac{\pi}{2},0)$:

1). The parameters of the mass matrix
$$
m_1 \approx \bar{m},~ m_2 \approx - \bar{m},~ a \approx \bar{m}/3, ~
b \approx - \frac{2\bar{m}}{3},~ c = 0
$$
give the TBM matrix  equals
\bea
m_{\nu} & \approx &  \bar{m} \left(
\begin{array}{ccc}
\frac{1}{3}  &  - \frac{2}{3}   & - \frac{2}{3} \\
 ... & - \frac{1}{6} &
- \frac{1}{6}    \\
 ... & ...  &  - \frac{1}{6} \\
\end{array} \right).\nonumber
\label{mtbm-inv2}
\eea
Now the elements of the $e-$ row dominate.\\

2). Lowest order corrections equal\\
\bea
x &=& \frac{\bar{m}}{3} \left( \frac{1}{\sqrt{2}} s_{13} e^{-i\delta}
 - 2 D_{23} \right),~~~
y =  -   \frac{\bar{m}}{3} \left(2\sqrt{2} s_{13} e^{i\delta}
- D_{23} \right), \nonumber\\
\Delta m_{ee} &=&   \bar{m} \left( 2 D_{12} - \frac{1}{3} s_{13}^2
\right),~~~~~~\Delta m_{\mu\tau} = -\bar{m}(D_{12}+\frac{1}{6}s_{13}^2).\nonumber
\eea

For the bf- (and $1 \sigma$) values of mixing parameters, we have
$x \sim - 0.25~ (- 0.5)$, $y \sim 0.2~ (0.3)$,
$\Delta m_{ee} \sim 0.2~ (0.3)$ in the units $10^{-2}$ eV.
Maximal values of the corrections can be achieved for $\delta = \pi$ see Table \ref{matr33max}.

The overall structure of the mass matrix does not change substantially.
Corrections correlate with zero order structure
being proportional to the same $\bar{m}$:
\be
m_{\nu} \approx   \frac{1}{3} \bar{m} \left(
\begin{array}{ccc}
1 + 6 D_{12} - s_{13}^2 & - 2 +
\frac{1}{\sqrt{2}} s_{13} e^{-i\delta} - 2 D_{23} & - 2
- \frac{1}{\sqrt{2}} s_{13} e^{-i\delta}
 + 2 D_{23}  \\
 ... & - \frac{1}{2} - 2\sqrt{2} s_{13} e^{i\delta} + D_{23} &
- \frac{1}{2}  \\
 ... & ...  &  - \frac{1}{2} + 2\sqrt{2} s_{13} e^{i\delta}
- D_{23}\\
\end{array} \right).\nonumber
\label{mtbm-inv3}
\ee

3). The parameters of violation of the TBM conditions:
Since the $e\mu-$ and $e\tau-$ elements are dominant
the relative corrections are small.
Indeed, the ``pole'' value of $s_{13}$ equals
\be
\tilde{s}_{13} \approx
\frac{2 s_{12} c_{12}}{(1 -  2 s_{12}^2)} \sim 3,\nonumber
\ee
and for the allowed range, $s_{13} \ll \tilde{s}_{13} $,
the TBM-violation is suppressed:
\be
\Delta_e \approx \frac{2s_{13}}{\tilde{s}_{13}}
\approx
\frac{2}{3} s_{13}.
\label{delta-mina}\nonumber
\ee

In contrast, since the original elements of the $\mu \tau-$ block are suppressed,
the relative corrections to $m_{\mu\mu}$ and $m_{\tau\tau}$ can be large, thus strongly
violating the 2nd TBM-condition. For $\Delta_{\mu \tau}$ we find the pole at
$\tilde{D}_{23} =\frac{1}{2}-s_{13}\tan 2\theta_{12}e^{i\delta}. $
At $1\sigma$ level,  $\tilde{D}_{23}\approx 0.09$, and one can achieve
$m_{\tau\tau}=0$,  as is shown in the Table~\ref{matr33}.

For $D_{23} = 0$ we have
\be
\Delta_{\mu \tau} \approx
\frac{4 s_{13} \sin 2 \theta_{12}}{\cos 2 \theta_{12} - 2s_{13}\sin
2\theta_{12}}.\nonumber
\ee
This dependence has a pole at $s_{13} = 0.5 \cot 2 \theta_{12}
\approx 0.17 - 0.20$,  at the maximal  allowed values of 1-3 - mixing.
Thus  $\Delta_{\mu \tau} \rightarrow \infty$  and violation of the
TBM structure  is strongly enhanced. According to (\ref{mtbm-inv3})
the $s_{13}-$ corrections are enhanced by additional factor $2\sqrt{2} \sim 3$.
For smaller values of $s_{13}$: $\Delta_{\mu \tau} \approx
4 s_{13} \tan 2 \theta_{12}$.\\

4). Properties of mass matrix:\\
\begin{itemize}

\item
the matrix can show ``inverted flavor hierarchy'' with
$m_{\tau\tau}$ being the smallest element;


\item
depending on $\delta$, $m_{\mu \mu} = 0$ or $m_{\tau \tau} = 0$
texture zero can be obtained at 1$\sigma$ level.

\end{itemize}
The basis corrections are
$\Delta m_{ee}^b=1.16 $,
$\Delta m_{\mu e}^b=-0.19 $,
$\Delta m_{e \tau}^b=0.09 $
and $\Delta m_{\mu \tau}^b=-0.22~$ in the units of $10^{-2}$ eV. Correction to $m_{ee}$ is large.

\subsection{Degenerate spectrum}

In the case of degenerate  spectrum,
$m_1 \approx m_2 \approx m_3 \approx m_0$, the structure of the mass matrix
depends strongly on values of both Majorana phases.

\noindent
A. $\phi_2 = \phi_3 = 0$, line $D(0,0)$.

1). Parameters of the mass matrix equal
\bea
 a \approx m_0,~~~~~
b \approx \epsilon_S \equiv \frac{\Delta m^2_{21}}{6 m_0},
&&
c = m_0 + \epsilon_A,~~~ ~~ a + b + c = 2 m_0 + \epsilon_A,\nonumber\\
~~
a + b - c &\approx& - \epsilon_A \equiv - \frac{\Delta m^2_{31}}{2 m_0}.
\nonumber
\eea
Numerically, for $m_0 = 0.2$ eV we find
$\epsilon_S = 6.7 \cdot 10^{-5}$ eV and
$\epsilon_A = 6.2 \cdot 10^{-3}$ eV.
The TBM mass matrix is very close to the unit matrix:
\be
m_{\nu}  \approx  \left(
\begin{array}{ccc}
m_0  &  \epsilon_S  & \epsilon_S  \\
 ... & m_0 + \frac{1}{2}\epsilon_A  & - \frac{1}{2}\epsilon_A   \\
 ... & ...  &  m_0 + \frac{1}{2}\epsilon_A \\
\end{array} \right) = m_0 I +
\left(
\begin{array}{ccc}
0  &  \epsilon_S  & \epsilon_S  \\
 ... &  \frac{1}{2} \epsilon_A  & - \frac{1}{2} \epsilon_A   \\
 ... & ...  & \frac{1}{2} \epsilon_A  \\
\end{array} \right) .
\label{mtbm-deg1}\nonumber
\ee
Furthermore, there is a strong hierarchy of the sub-leading
(off-diagonal) elements.\\

2). The lowest order corrections:
 Neglecting  terms proportional to $\epsilon_S$ we find
\be
x =  - \frac{1}{\sqrt{2}} s_{13}
[(m_0 + \epsilon_A) e^{-i\delta} - m_0 e^{i\delta}],~~~y = - \epsilon_A D_{23},~~~
\Delta m_{ee} =   s_{13}^2
\left[(m_0 + \epsilon_A) e^{-2i\delta} - m_0 \right],\nonumber
\ee
and the size of corrections strongly depends on $\delta$:
\be
x \approx
\left\{
\begin{array}{ll}
- \frac{1}{\sqrt{2}} s_{13} \epsilon_A & \delta = 0 \cr
i\sqrt{2} s_{13} m_0 & \delta = \pi/2
\end{array}
\right. ~.\nonumber
\ee
 For $\delta = \pi/2$ the correction $\Delta m_{ee}$ is maximal:
$\Delta m_{ee} =  - 2 m_0 s_{13}^2$.\\

3). Parameters of violation of the TBM-conditions:
For $\phi_2  = 0$,  the $e\mu-$ and $e\tau- $ elements are very
small,  so that they can be canceled at very small $\tilde{s}_{13}$.
Indeed,
\be
\tilde{s}_{13} \approx s_{12} c_{12}
\frac{\Delta m^2_{21}}{\Delta m_{31}^2}
= 0.015.\nonumber
\ee
Correspondingly, the singularity and the peak move to small values of
$s_{13}$.

The pole value $\tilde{D}_{23} \approx-1/s_{13}^2e^{2i\delta}\rightarrow \infty$,
so that
\be
\Delta_{\mu \tau} \approx D_{23} \frac{\Delta m_{31}^2}{2 m_0^2}~\nonumber
\ee
turns out to be strongly suppressed as a consequence of
dominance of the $\mu\mu-$ and $\tau\tau-$ elements.
For $s_{13} \neq 0$ and $D_{23} = 0$:
$\Delta_{\mu \tau} \approx s_{13} \sin 2 \theta_{12}
\Delta m_{21}^2/2 m_{0}^2$,
and for $m_{0} = 0.2$ eV  the breaking
is very strongly suppressed due to smallness of $b$:
$\Delta_{\mu \tau} \sim - 10^{-3} s_{13}$.\\

4). Properties of the mass matrix:
\begin{itemize}
\item
Corrections do not affect the dominant elements but can change
completely the sub-dominant structure.

\item
The only significant change in
neutrino mass matrix is
violation of equality of $m_{e\mu}$ and $m_{e\tau}$:
$\Delta_e \approx 2$
can be achieved, which  corresponds to $m_{e\tau}\simeq -m_{e\mu}$:
\bea
m_{\nu} \simeq m_0~I +
\left(
  \begin{array}{ccc}
   \Delta m_{ee}  &  x & -x \\
    ... & y &  - \frac{1}{2} \epsilon_A  + \Delta m_{\mu \tau} \\
    ... & ... & -y \\
  \end{array}
\right).\label{massdeg}\nonumber
\eea
Notice that due to corrections the elements of the second
sub-dominant matrix in (\ref{massdeg}) can be of the same order:
$|x| \sim |y| \sim  \epsilon_A $, or can obey certain symmetry.

\item
Since $\tilde{s}_{13}$ is very small all special mass
relations indicated in the Table 3, including texture zeros,
can be achieved.

\end{itemize}
The basis corrections equal
$\Delta m_{ee}^b=0.01,~
\Delta m_{\mu e}^b=0.06 ,~
\Delta m_{e \tau}^b=0.06 $
and $\Delta m_{\mu \tau}^b=-0.06$ (in the units $10^{-2}$ eV). They are of the order of the TBM deviation corrections for $e\mu-$ and $e\tau-$ elements.

\noindent
B. $\phi_2 = 0$, $\phi_3 =\pi/2$; line $D(0, \frac{\pi}{2})$.\\

1). The parameters of TBM mass matrix
 $$a \approx m_0, ~~~b \approx \epsilon_S,~~~
c = - m_0 - \epsilon_A,~~~
a + b + c \approx - \epsilon_A,~~~
a + b - c \approx 2 m_0 + \epsilon_A,~~~ a - c = 2 m_0 + \epsilon_A$$
give TBM matrix
\be
m_{\nu}  \approx  \left(
\begin{array}{ccc}
m_0  &  \epsilon_S  & \epsilon_S  \\
 ... & -\frac{1}{2}\epsilon_A  & m_0 + \frac{1}{2}\epsilon_A   \\
 ... & ...  &  - \frac{1}{2}\epsilon_A \\
\end{array} \right) = m_0 T +
\left(
\begin{array}{ccc}
0  &  \epsilon_S  & \epsilon_S  \\
 ... &  - \frac{1}{2} \epsilon_A  &  \frac{1}{2} \epsilon_A   \\
 ... & ...  & - \frac{1}{2} \epsilon_A  \\
\end{array} \right),
\label{mtbm-deg1}\nonumber
\ee
where $T$ is the ``triangle'' matrix with the only non-zero elements
$m_{ee} = m_{\mu \tau} = m_{\tau \mu}$.
The elements of the matrix are strongly hierarchical.\\

2). The lowest order corrections equal
 \be
x = \frac{1}{\sqrt{2}} s_{13}
[(m_0 + \epsilon_A) e^{-i\delta} + m_0 e^{i\delta}], ~~~
y = - 2 m_0 D_{23},~~~ \Delta m_{ee} =  2 s_{13}^2 m_0.\nonumber
\ee
The largest deviation is for $\delta = 0$:
$x \approx  \sqrt{2} s_{13} m_0$.\\

3). The parameters of  violation of the TBM conditions:\\
\be
\tilde{s}_{13} \approx s_{12} c_{12}
\frac{\Delta m^2_{21}}{4 m_{0}^2} = 2.5 \cdot 10^{-4}\nonumber
\ee
for $m_0 = 0.2$ eV.  The reason for this smallness is that the original
elements of the $e$-row are very small.
For the allowed values of $s_{13}$
the maximal TBM-violation, $\Delta_e \approx 2$,
can be nearly achieved and all special mass relations
of the Table 3 can be realized.

The $\mu\mu-$ and $\tau\tau-$ elements are strongly suppressed and
corrections dominate. The pole of $\Delta_{\mu \tau}$, which
corresponds to $m_{\tau\tau} = 0$, is at
\be
\tilde{D}_{23} \approx
- \frac{\Delta m_{31}^2}{8 m_0^2}~ = - 0.008
\nonumber
\ee%
and is achieved for the negative values of $D_{23}$ ($\theta_{23} >
\pi/4$). Then from (\ref{d-mut1}) we obtain
\be
\Delta_{\mu \tau}
\approx 2 \frac{D_{23}}{D_{23} + \frac{\Delta m_{31}^2}{8 m_0^2} (1
+ 2 D_{23})} ~. \label{d-mutau2}\nonumber
\ee%
Numerically, we have
$\tilde{D}_{23} = - (0.03, ~ 0.0075, ~ 0.0033)$ for $m_0 = (0.1,
0.2, 0.3)$ eV correspondingly. This pole value is well within the
$1\sigma$ allowed range for $D_{23}$, so, all special mass relations
of the Table 4 can be obtained. 
In particular, for positive $D_{23}$ at $D_{23} \approx - \tilde{D}_{23}$, we have
$\Delta_{\mu \tau} = 1$ which corresponds to $m_{\mu \mu} = 0$.
For $|D_{23}| \gg  |\tilde{D}_{23}|$
$\Delta_{\mu \tau} \rightarrow 2 $
independently of the sign of $D_{23}$. This value of $\Delta_{\mu\tau}$ corresponds to
$m_{\mu \mu} \simeq -m_{\tau \tau}$. Notice that for
$m_0 = 0.2$ eV and $1\sigma$ allowed
$D_{23}$ the ratio $D_{23}/ \tilde{D}_{23} \approx 12$,
so that the limit can be
realized with  a good accuracy.

For non-zero 1-3 mixing but $D_{23} = 0$:
\be
\Delta_{\mu \tau} \approx  2 s_{13} \sin 2 \theta_{12}
\frac{\Delta m_{21}^2}{4 m_{0}^2 s_{13}^2 +  \Delta m_{21}^2 c_{12}^2}.\nonumber
\ee
For $s_{13} >> c_{12} \sqrt{\Delta_{21}^2}/2m_0 \sim 10^{-2}$ we have
\be
\Delta_{\mu \tau} \approx  \frac{\sin 2 \theta_{12}}{s_{13}}
\frac{\Delta m_{21}^2}{2 m_{0}^2}~.\nonumber
\ee
The deviation increases with decrease of $s_{13}$, however, even
in maximum $\Delta_{\mu \tau}$ does not exceed $0.03$.\\

4). Properties of the mass matrix:
\begin{itemize}

\item
With corrections  the neutrino mass matrix
takes the following form
\bea
m_{\nu} \approx
m_0 \left(\begin{array}{ccc}
1 &  \sqrt{2} s_{13}  & - \sqrt{2} s_{13} \\
... & 2D_{23} - \frac{1}{2m_0} \epsilon_A & 1 + \frac{1}{2m_0} \epsilon_A \\
... & ... & - 2D_{23} - \frac{1}{2m_0} \epsilon_A
\end{array}
\right).
\label{p10p2pi/2}\nonumber
\eea
Here two TBM-conditions are maximally broken,
however $x \ll a$.
Corrections have completely different symmetry from that of the dominant
block which has the ``triangle'' form.

\item
Corrections to the dominant triangle structure can be all of the same order
and of the size of Cabibbo angle with respect to the dominant structure:
$$
m_{\nu} = m_0[T + 0.2 D],
$$
where $D$ is the ``democratic'' matrix or
matrix with elements of the same order.

\item
Since both $\tilde{s}_{13}$ and $\tilde{D}_{23}$ are very small,
special relations for elements of the e-line and $\mu\tau-$ block
can be satisfied simultaneously.

\end{itemize}
The basis corrections are
$\Delta m_{ee}^b=-0.81$,
$\Delta m_{\mu e}^b=-4.0 $,
$\Delta m_{e \tau}^b=-4.1 $ and
$\Delta m_{\mu \tau}^b=4.1 $ (in the units $10^{-2}$ eV). The corrections to the $e\mu$ and $e\tau$ elements are of the order of corrections due to the TBM mixing deviations.

\noindent
C. $\phi_2 = \pi/2$ and $\phi_3 = 0$; line $D(\frac{\pi}{2},0)$.\\

1). The parameters of the mass matrix
$$
m_2 \approx - m_0, ~~~a = \frac{m_0}{3},~~~
b = - \frac{2m_0}{3}, ~~~c =  m_0,~~~
a + b + c \approx \frac{ 2m_0}{3},~~~ a + b - c \approx  - \frac{4 m_0}{3}
$$
lead to TBM-matrix
\be
m_{\nu}  \approx  m_0 \left(
\begin{array}{ccc}
\frac{1}{3}  &   - \frac{2}{3}  & - \frac{2}{3}   \\
 ... &  \frac{1}{3}   &   - \frac{2}{3}   \\
 ... & ...  &  \frac{1}{3}  \\
\end{array} \right) = m_0 I - \frac{2}{3} m_0 D,
\label{mtbm-deg2}
\ee
where $D$ is the democratic matrix.\\

2). The lowest order corrections equal
\bea
x & = & - \frac{1}{\sqrt{2}} s_{13}
(e^{-i\delta} - \frac{1}{3} e^{i\delta})m_0 - \frac{2}{3} D_{23} m_0,
\nonumber\\
y & = & - \frac{2}{3}m_0 \left(2 D_{23} + \sqrt{2} s_{13}
e^{i\delta} \right),
\nonumber\\
\Delta m_{ee} & = & 2 m_0 \left(D_{12} + \frac{1}{3}s_{13}^2 \right),
~~~~~\Delta m_{\mu\tau} = m_0 \left(-D_{12} + \frac{1}{3}s_{13}^2\right).
\nonumber
\eea
Relative corrections are enhanced because the elements of
original matrix are suppressed by numerical factors.
Corrections equal   $50\%, ~ (100\%)$ for $y$,
$20\%~ (30\%)$  for $x$ and $10 (20\%)$ for the $ee-$element.\\

3). The violation of the TBM conditions:
The original elements
$e\mu-$ and $e\tau-$ are large and $s_{13}$ produces
relatively small effect. The pole value equals
$\tilde{s}_{13} \approx \cot \theta_{12} > 1$,
as a result,  for the allowed values of $s_{13}$
the breaking parameter is suppressed:
\be
\Delta_e = \frac{2s_{13}}{
\cot \theta_{12} e^{i\tilde{\phi}} - s_{13}}
\approx 2s_{13} \tan  \theta_{12}.
\label{d-mina}
\ee
The pole of $\Delta_{\mu\tau}$ is given by
\be
\tilde{D}_{23} =
- \frac{1}{2} \tan^2\theta_{12}
\left(1 + 2 \cot\theta_{12}  s_{13}e^{i\delta}\right)
\approx - \frac{1}{4}\left(1 + 2\sqrt{2} s_{13}e^{i\delta}\right).\nonumber
\ee
Consequently,
$\tilde{D}_{23}\gg D_{23}^{max}$ for $\delta=0$. The minimal value of
$\tilde{D}_{23}$ is realized at $\delta = \pi$ and maximal possible
$s_{13}$: $\tilde{D}_{23} \sim 0.1$ for which $\tilde{D}_{23} \approx
D_{23}^{max}$. Therefore one can reach the pole and
$m_{\tau\tau} \approx 0$. Since $\phi_2=\pi/2$, the parameter
$\beta$ is not suppressed and become important:
$$
\beta=\frac{s_{13}\sin2\theta_{12}e^{i\delta}}{2(1+ \cos2\theta_{12})}~.
$$
Texture zero $m_{\mu\mu}=0$ is realized
if $D_{23}=-(\tilde D_{23}+\beta)$. For $\delta = 0$ we obtain
at the $1\sigma$ level  $-(\tilde D_{23}+\beta)\approx  0.085$
which is close to the $1\sigma$ allowed value of $D_{23}$.
Consequently, $m_{\mu\mu}\approx 0$ can be achieved at $1\sigma$
level, as can be seen in  the Table \ref{matr33}.

For $s_{13} = 0$ we have
\be
\Delta_{\mu \tau} \approx
4 \frac{D_{23}}{\tan^2 \theta_{12} + 2 D_{23}}~.
\label{d-mutau3a}
\ee
For negative $D_{23}$ the deviation can be substantially enhanced
($\sim 12 D_{23}$), still the pole
is not realized.

The 1-3 mixing effect (for $D_{23} = 0$) on violation
of the second TBM condition is given by
\be
\Delta_{\mu \tau} \approx 4 s_{13} \frac{c_{12}}{s_{13} + 2 s_{13} c_{12}}
\approx 4 s_{13} \cot \theta_{12}~.\nonumber
\ee
Here correction is enhanced by the factor
$4 \cot \theta_{12} \sim 6$, so that for  maximal allowed 1-3 mixings
we obtain $\Delta_{\mu \tau} \approx 1$.\\

4). Properties of the mass matrix:
 \begin{itemize}
\item
the $e\mu-$ and $e\tau-$ elements can differ by
$50 - 60 \%$,

\item
texture zeros $m_{\mu \mu} = 0$ or $m_{\tau \tau} = 0$
can be achieved;

\item
the equalities
$m_{ee} \approx - m_{e\tau}$, $m_{\mu \tau} = m_{\tau\tau}$ are possible;

\item
the matrix may have rather random ``anarchical'' character;

\item
at $1\sigma$ level the structure of the matrix can change strongly,
and the TBM conditions can be strongly broken.

\end{itemize}
The basis corrections are
$\Delta m_{ee}^b=5.2~$,
$\Delta m_{\mu e}^b=1.1 ~$,
$\Delta  m_{e \tau}^b=-2.4 ~$
and $\Delta m_{\mu \tau}^b=2.98 ~$ (in the units $10^{-2}$ eV). They are of the order of TBM corrections for $e\mu-$ and $e\tau-$ elements and large for $ee-$ elements.

\noindent
D. $\phi_2 =  \phi_3 =\frac{\pi}{2}$, line ($D(\frac{\pi}{2},\frac{\pi}{2})$):\\

1). Parameters of the mass matrix
  \bea
  m_1 &=& m_0,~~~ m_2 \approx - m_0 - \epsilon_S,~~~
m_3 \approx - m_0 - \epsilon_A,~~~
a = \frac{m_0}{3},\nonumber\\ b &=& - \frac{2m_0}{3}, ~~~~c = -m_0,~~~~
a + b + c \approx  -\frac{4m_0}{3}, ~~~~a + b - c \approx
\frac{2m_0}{3}\nonumber
\eea

give the TBM-mass matrix
\be
m_{\nu}  \approx  m_0 \left(
\begin{array}{ccc}
\frac{1}{3}  &   - \frac{2}{3}  & - \frac{2}{3}   \\
 ... &  - \frac{2}{3}   &    \frac{1}{3}   \\
 ... & ...  &  \frac{2}{3}  \\
\end{array} \right) = - m_0 V_1 =  m_0 T - \frac{2}{3} m_0 D,
\label{mtbm-deg2}
\ee
where $D$ is the democratic matrix.
This matrix differs from the one in the previous case by permutation
in the $\mu \tau-$block. It is proportional to the symmetry matrix $V_1$.\\

2). The lowest order corrections equal
\bea
x & = &  \frac{1}{\sqrt{2}} s_{13} (e^{-i\delta}
+ \frac{1}{3} e^{i\delta})m_0 - \frac{2}{3} D_{23} m_0,
\nonumber\\
y & = &  \frac{2}{3}m_0 \left(D_{23} + \sqrt{2} s_{13} e^{i\delta}\right),
\nonumber\\
\Delta m_{ee} & = & 2 m_0 \left(D_{12} - \frac{2}{3}s_{13}^2 \right).\nonumber
\eea
The deviation $x$ is enhanced if $\delta = \pi$:
\be
x = - \frac{2}{3}m_0 (\sqrt{2} s_{13} + D_{23}).\nonumber
\ee
In this case $y  =   \frac{2}{3}m_0 (D_{23} - \sqrt{2} s_{13})$.
All the elements of the TBM matrix are of the same order and just differ
by factor 2. The elements of the $e$-raw
and $\mu\tau-$ block affected by the corrections are
large, and therefore effect of corrections is relatively small:
for the bf-values and
$1\sigma$ we have   $12\%, ~ (25\%)$ for $y$ ($\mu \tau-$block),
and $25\%~ (45\%)$  for $x$ and $8 (10\%)$ for the $ee-$element.\\

3). The parameters of violation of the TBM conditions:
Now $\tilde{s}_{13} \approx \tan \theta_{12}$,
so that
\be
\Delta_e = \frac{2s_{13}}{
\tan \theta_{12} e^{i\tilde{\phi}} - s_{13}}
\approx 2s_{13} \cot \theta_{12}.
\label{delta-e3a}
\ee
The violation parameter $\Delta_{\mu \tau}$ equals
\be
\Delta_{\mu \tau} \approx
4 D_{23} \frac{D_{23}}{\cot^2\theta_{12} + 2 D_{23}}~.
\label{d-mutau4a}\nonumber
\ee
Here enhancement is weaker than in the previous case.
For $D_{23} = 0$ we have
\be
\Delta_{\mu \tau} \approx  4 s_{13} \frac{s_{12}}{c_{12} - 2 s_{13} s_{12}}
\approx  4 s_{13} \tan \theta_{12}~.\nonumber
\ee
Since
$$
\tilde D_{23}= - \frac{c^2_{12} -
\sin\theta_{12}s_{13}e^{i\delta}}{2s^2_{12}},~~~ ~|\tilde D_{23}|\gg D_{23}^{max},
$$
no zeros can be obtained
in the $\mu\tau$-block.\\

4). Properties of mass matrix:
It may have the form
\bea
m_{\nu} &\approx&
\left(
  \begin{array}{ccc}
    a & y & z \\
    y & z & a \\
    z & a & y \\
  \end{array}
\right) + \delta m_\nu \nonumber\\ &=&
m_0 \left(
  \begin{array}{ccc}
    \frac{1}{3}  & - \frac{2}{3}(1+D_{23}) +
\frac{\sqrt{2}}{3} \sin 2 \theta_{13}  & - \frac{2}{3}(1-D_{23}) - \frac{\sqrt{2}}{3}
\sin 2 \theta_{13}  \\
  ... &  - \frac{2}{3} + \frac{y}{m_0}  &  \frac{1}{3}  \\
  ... & ... &  - \frac{2}{3} - \frac{y}{m_0} \\
  \end{array}
\right).
\label{p1p2pi/2}\nonumber
\eea
This matrix has approximate cyclic symmetry and
the element of second diagonal are equal
(see also the line $D\left(\frac{\pi}{2},\frac{\pi}{2}\right)$ in the Table
\ref{matr33}).


If $\delta = 0$, then $y$ is enhanced:
$y  = \frac{2}{3}m_0 (D_{23} + \sqrt{2} s_{13})$, the
two contributions sum up. At the same time in $x$
the two contributions partially cancel each other.
The elements of mass matrix  have rather random spread within factor
3, without clear structure. The TBM conditions are broken by
$O(1)$ factors.


The basis corrections for $s_b=0.2$ equal
$\Delta m_{ee}^b=4.4~$,
$\Delta m_{\mu e}^b=-2.9 $,
$\Delta  m_{e \tau}^b=-1.6 $  and $\Delta m_{\mu \tau}^b=1.1
$ (in the units $~10^{-2}$ eV). They are large for the $ee-$ element and significant for the $e\mu-$ and $e\tau-$ elements.

\section{Deviations from TBM and flavor symmetry}

Using results of the previous sections we
will  consider implications of the mass
matrices with deviations from TBM
structure for the flavor symmetries.

Recall that the  TBM as well as other flavor structures
could be immediate consequence of symmetry,
if e.g. (i)  single mechanism of neutrino mass
generation dominates  and various corrections
are negligible;  (ii) Higgses are flavorless,
so that the  problem  of VEV alignment does not exist.
In this case one needs to adjust the Yukawa coupling constants only.
It can be shown, however, that flavor symmetry, should be broken
to explain TBM. The flavor structures which can be obtained in this
scenario do not reproduce TBM but they
can serve as the dominant structures of the mass matrix.

The operator responsible for the Majorana neutrino masses has the form
\be
L = h_{ij} L_i L_j X,\nonumber
\ee
where, in general, the ``Higgs factor'' $X$ is
some combination of  the Higgs fields.

\subsection{Deviations from TBM and new flavor symmetries?}

Do neutrino mass matrices with deviations from TBM have
some new symmetry which differs from
the TBM symmetry? Here we briefly mark some possibilities,
their detailed realizations will be presented elsewhere~\cite{AbSm}.

As we have established in the previous sections the
corrections can lead to new  equalities between the matrix elements. In particular,
\be
m_{e\mu} \approx - m_{e\tau},
\label{eq-em}
\ee
as well as $m_{\mu\mu} = - m_{\tau\tau}$,
$m_{e\mu} =  m_{\mu\mu}$, $m_{e \mu} = n m_{e \tau}$ with, e.g.,
$n = 2, 1/2$, {\it etc.}.
Let us consider implications of the  equality (\ref{eq-em}).

If $b = m_{e\mu}^0$ and the deviation $D_{23}$ are very small, then
the correction $x$ dominates, $y$ is negligible, and furthermore,
the corrections of the order $s_{13} D_{23}$, which contribute to
$m_{e \mu}$  and $m_{e \tau}$ equally, are also small.
In this case
the mass matrix has the following
approximate form:
\be
\left(
\begin{array}{ccc}
      a & - x & x \\
      ... & \frac{1}{2}(a+c) & \frac{1}{2}(a-c) \\
      ... & ... & \frac{1}{2}(a+c) \\
    \end{array}
  \right).
\label{sym-mat}\nonumber
\ee
The conditions for this form of matrix are realized in the cases of spectra with
quasi-degenerate first and second states: $m_1 \approx m_2$ and
$\phi_2 = 0$: $PD(0,0)$, $PD\left(0, \frac{\pi}{2}\right)$,
$IH\left(0, 0 \right)$,
$D\left(0, 0 \right)$,  $D\left(0, \frac{\pi}{2}\right).$
(Notice that in the Table \ref{matr33} the
examples of matrices correspond to maximal allowed value of $D_{23}$,
so that correction $s_{13}D_{23}$ leads to violation of equality
(\ref{eq-em})).

In the case of inverted mass hierarchy, $IH\left(0, 0\right)$,
also $c\simeq 0$.
The  matrix (\ref{sym-mat}) is invariant under the transformation
$$
V_2^{\prime} = \left(\begin{array}{ccc}
1 & 0 & 0 \\
0 & 0 & -1 \\
0 & -1 & 0 \\
\end{array}
\right)
$$
which is one the generators of $S_4$.
This is new residual symmetry, since the first
TBM conditions is broken by the  1-3 mixing.
As we have mentioned before, the equality (\ref{eq-em})
and symmetry with respect to $V_2$ (\ref{v1v2}) can be restored
by redefinition: $\nu_\mu \rightarrow - \nu_\mu$.
In this case the $\mu \tau-$element changes the sign:
$\frac{1}{2}(a-c) \rightarrow - \frac{1}{2}(a-c)$ and the third
TBM condition turns out to be broken:
$\Sigma_L - \Sigma_R = a - c +x$.
 Now the matrix is invariant with respect to
$V_2$ but not $V_1$.\\

In contrast to the TBM matrix, the matrices with
deviations can contain texture zeros \cite{texturezero} and agree with
observed  neutrino masses.
Interesting examples, which can testify for certain symmetries,
follow:

1. The texture zeros $m_{e\mu} = 0$ or $m_{e\tau} = 0$ can be achieved
in the  cases of normal mass hierarchy: $NH(0,0)$, $NH(0,
\frac{\pi}{2})$,
partial degeneracy: $PD(0, 0)$, $PD(0, \frac{\pi}{2})$,
inverted hierarchy: $IH(0, 0)$, degenerate spectrum:
$D(0,0)$, $D(0, \frac{\pi}{2})$. In all these cases the original
elements of the $e$-line are small.

2. The texture zeros $m_{\mu\mu} = 0$ or $m_{\tau\tau} = 0$ can be
obtained in the
cases of inverted mass hierarchy $IH(\frac{\pi}{2},0)$, and degenerate
spectrum $D(0, \frac{\pi}{2})$. The condition for that is
$m_{\mu\mu}^0 = m_{\tau\tau}^0 \ll m_{\mu\tau}^0$.

3. Matrices with two texture zeros become allowed:
various combinations of zeros in the $e-$line and
$\mu\tau-$block (indicated above) can be obtained for the degenerate spectrum $D(0, \frac{\pi}{2})$.
In particular, in the case of very small 1-3 mixing
($s_{13}=0.001$), $D_{23}\sim 0.032$, $\delta=0$ and $m_1=0.09$ eV,
all the elements of the second diagonal can be
zero, $m_{\mu\mu} = m_{e\tau} = 0$. By changing the value
of $\delta$ and the sign of $D_{23}$, we can change the positions
of ``zeros''. If $D_{23}\sim -0.032$, the two texture zeros
$m_{\tau\tau}=m_{e\tau}=0$ are achieved. If $\delta=\pi$, we get
$m_{\mu\mu} = m_{e\mu} = 0$, so that the 1-2 mixing is induced.
If $D_{23}\sim -0.032$ and $\delta=\pi$, we get $m_{\tau\tau}=m_{e\mu}=0$.

4. An interesting possibility is the matrix with two texture zeros:
$m_{e\mu} = m_{ee} = 0$ which can be achieved in the case of normal mass
hierarchy with $m_1 \sim 0.0031$ eV at the best fit values of the mixing angles and $(\phi_2, \phi_3, \delta)=(\pi/2, 0, \pi)$.
This is signature of yet another class of
underlying symmetries.

5. Also the Fritzsch-type matrix with $m_{ee} = 0$, $m_{e\tau} = 0$  and
relatively small $m_{\mu\mu}$ can be realized in the case of normal mass hierarchy and $m_1 \sim 0.0035$ eV at the best fit values of the mixing angles and $(\phi_2, \phi_3, \delta)=(\pi/2, 0, 0)$.

\subsection{Two-component structure of the mass matrix}

In a number of cases the neutrino mass matrix has
strongly hierarchical structure with
large elements forming the dominant block
and small sub-dominant elements.
This may indicate that the mass matrix has a two-component structure
\be
m_{\nu} = M_d + \mu_{s},
\label{sum}
\ee
where $M_d$ and $\mu_s$ are the dominant and sub-dominant contributions.
The matrices $M_d$ and $\mu_s$ may have different origins and  different
symmetries, the sub-dominant matrix
$\mu_{s}$ may appear as a  result of breaking of symmetry of
$M_d$, and symmetry can be completely broken in $\mu_s$.

As we have shown the relative corrections to the dominant block
elements  are of the order $30\%$, whereas corrections to the
sub-dominant elements can be much larger than the original elements.
Therefore if the mass matrix, indeed, has two different contributions,
the corrections can completely change the structure and possible
symmetries of the sub-dominant matrix.
There are different scenarios for (\ref{sum}).
The dominant $M_d$ can be a consequence of unbroken symmetry,
whereas the sub-dominant block appears as a result of symmetry
breaking.

Here we briefly consider possible symmetries which lead to various
dominant structures:

1.  The $\mu-\tau-$ dominant block (the case of normal mass hierarchy)
has, e.g., the $U(1)-$symmetry with the charge prescriptions
$L(\nu_e) = 1$,  $L (\nu_\mu) = L (\nu_\tau) = 0$ \cite{le-lmu-ltau}.

2. The matrix with the dominant block,
which consists of the $\mu\mu-$, $\tau\tau-$, $\mu\tau-$
and $ee-$elements is realized in the case of partially
degenerate spectrum $PD(0,0)$. Is is invariant under
\be
\nu_e \rightarrow \nu_e, ~~ \nu_\mu \rightarrow - \nu_\mu, ~~
\nu_\tau \rightarrow - \nu_\tau .\nonumber
\ee
Clearly this symmetry cannot be exact symmetry of the whole
Lagrangian, but  it can appear as a residual summery for
neutrino Yukawa couplings.


3. The matrix proportional to the unit matrix, $M_d = m_0 I $,
is the dominant structure for the degenerate spectrum $D(0,0)$.
It can be a consequence of various discrete  and continuous symmetries.
Suppose the lepton doublets, $L_i$, form triplet of some symmetry group
$G_f$:  $L\sim {\bf 3}$,  and Higgses are flavorless. Then
to get invariant combination ${\bf 3} \times {\bf 3} \sim {\bf 1}$
the group $G_f$ should be $SO(3)$ or some its subgroup.
The smallest group with irreducible representation
${\bf 3}$ is $A_4$ and the invariant combination
$L_i L_i$  produces the required unit matrix.

Suppose  the Higgs factor $X$ is singlet of symmetry group but
not invariant, e.g. $X \sim {\bf 1^\prime}$ or $X \sim {\bf 1''}$
of $A_4$, then  $L_i L_j$  should transform as ${\bf 1''}$  and
${\bf 1^\prime}$ correspondingly. These combinations produce either zero
mass   (because of the antisymmetric nature of couplings)
or the matrix proportional to the diagonal phase matrix:
$m_\nu = m_0 {\rm diag} (1, e^{2i \pi/3}, e^{4 i\pi /3})$.


4. The triangle matrix $M_d = m_0 T$ is the dominant structure in the
case of degenerate spectrum $D(0, \frac{\pi}{2})$. This structure
can be a consequence of discrete or continuous symmetries,
as in the previous case. In particular, the $A_4$ model with
triplet $L_i$, in the complex representation
leads to the triangle form.


Also the triangle dominant structure with $m_{ee} \neq m_{\mu\tau}$
is possible in the case of deviation from TBM.
This structure can be produced in models
where $\nu_\mu$ and $\nu_\tau$ form a doublet of some (discrete) symmetry
group:
$L_1 \sim {\bf 1}$ and $\tilde{L}=\left(
                                    L_2,L_3
                                  \right)^T
\sim {\bf 2}$.  The neutrino mass matrix is diagonal
for real representation and  of triangle form  for  complex representation.
Such a situation can be realized
in the case of $S_3$ group and its further embedding like $S_4$, {\it  etc.}.

If the lepton doublets transform as  singlets  of the symmetry group:
e.g., $L_1 \sim {\bf 1}$, $L_2\sim {\bf 1'}, L_3 \sim {\bf 1''}$
(and $X \sim {\bf 1}$)
the neutrino mass matrix is of the triangle form:
\bea
m_{\nu}=v
\left(
  \begin{array}{ccc}
    h_{11} & 0 & 0 \\
    0 & 0 & h_{23} \\
    0 & h_{23} & 0 \\
  \end{array}
\right),\nonumber
\eea
where $h_{11}$ and $h_{23}$ can be of the same order.

A possibility to get some flavor structures
immediately from symmetry is  to use a single Yukawa  coupling, but
$X$  having  non-trivial flavor structure.
If $L\sim 3$ and $X \sim {\bf 3}$, the neutrino mass matrix equals
\bea
m_\nu=h \left(
          \begin{array}{ccc}
            v_1 & v_3 & v_2 \\
            v_3 & v_2 & v_1 \\
            v_2 & v_1 & v_3 \\
          \end{array}
        \right).
\label{recstruc}\nonumber
        \eea%
The TBM form can be achieved if $v_2=v_3$
but in this case $| m_1|=| m_2|$.
With possible deviations from TBM we can easily reproduce
the required structure (\ref{recstruc}) with $m_1\neq m_2$.
It appears in the case of
degenerate spectrum $D\left(\frac{\pi}{2}, \frac{\pi}{2}\right)$
(see eq. (\ref{p1p2pi/2})).
The problem is reduced now to VEV alignment:
$v_1 \approx 1/3$, $v_2 \approx -2/3 -x$, $v_2 \approx -2/3 + x$
and $x = - y$.

\subsection{No-symmetry case}

1. {\it Anarchical matrix}  with random values of elements
\cite{anarchy} is an extreme case. Matrix of this type appears for
certain intervals of CP-phases  in the cases of partial degeneracy
or degenerate spectra: $PD\left(\frac{\pi}{2},0\right)$,
$D\left(\frac{\pi}{2}, \frac{\pi}{2}\right)$, $D\left(\frac{\pi}{2},
0 \right)$ when the original TBM mass matrix has no or weak
hierarchy of elements. In these the ``random'' mass matrix leads accidentally to
strong degeneracy mass eigenstates. This implies fine tunning unless
certain new symmetry is introduced. Alternatively, this can imply that the mixing comes from the charged lepton sector whereas neutrino mass matrix has diagonal quasi-degenerate form and obey certain symmetry.

There are various possible origins of the anarchical structure, for
instance,  the see-saw mechanism with many ($n \gg 3$) right-handed
neutrinos. Another possibility is when two different and independent mechanisms give
comparable contributions  to the mass matrix. Each of these
contributions separately may have rather regular structure.

\subsection{Matrices with  flavor alignment}
There are two possibilities:

1). Normal flavor alignment. In the
case of normal mass hierarchy with $m_1 \neq 0$ the corrections due
to deviations from TBM as well as  basis corrections can wash out
sharp difference between the elements of the $\mu \tau-$ block and
$e-$ line. As a result one obtains a gradual decrease of size of
elements from $m_{\tau \tau}$ to $m_{ee}$.

2). In the case of
inverted mass hierarchy (see $IH(\frac{\pi}{2},0)$ ) the corrections
can produce an inverse flavor hierarchy when the values of matrix
elements increase with moving from $\tau-$ to $\mu-$ flavors.

These possibilities may indicate some perturbative origins and a
kind of the Froggatt-Nielsen mechanism \cite{fn mechanism} with large expansion parameter.

\section{Conclusion}
Is the TBM mixing accidental? The question is reduced, essentially,
to the question  whether this mixing immediately follows from some
(broken) symmetry or other principle, or it appears as a result of
many-step construction, and fixing various parameters
by introduction of additional
symmetries and structures.

The symmetry is formulated at the level of
mass matrix. Therefore if the data imply very specific mass
matrix with small deviations from the TBM form, we can
say that TBM is not accidental. We find the opposite: very strong
deviations of $m_\nu$ from $m_{TBM}$ and strong violations of the
TBM conditions (immediate manifestation of the symmetry) are
allowed.
This can be considered as an indication that TBM is accidental.
We find that large variety of the mass matrices with deviations from TBM
explain experimental data.

Strong deviations of $m_\nu$ from $m_{TBM}$ opens up a possibility
of the some alternative approaches to explain the data. Namely, some other
symmetry (which differs from the TBM symmetry) or other
principle can be involved. For instance, matrices  with texture zeros
are allowed which indicates, e.g. $U(1)$ underlying symmetry. Also
matrices with different relations between the elements are
possible, which testify for yet another class of symmetries.


We show that the mass matrix may show no trace of symmetry having random
values of elements. However, this corresponds to the
quasi-degenerate spectrum which implies another way to explain
the data.
In some cases the matrix has certain flavor alignment: gradual
change  of values of matrix elements from $m_{ee}$ to $m_{\tau\tau}$.

For certain ranges of masses and CP-phases the mass matrix has
structure with strong hierarchy between matrix elements: dominant
and sub-dominant ones.
We find that corrections can change  the
dominant elements by factors $O(1)$ and be much larger than the
sub-dominant elements. This may support the idea of
two-component structure of the mass matrix when the dominant block
has certain (unbroken) flavor symmetry and appears at the lowest
renormalizable level, whereas the sub-dominant structures can be
result of symmetry breaking by, e.g., high order
operators with flavon fields.

If it turns out that these new approaches lead to simpler and more
straightforward explanation of the data, the TBM symmetry approach
will be disfavored.


The 1-3 mixing leads to the most strong corrections. So,
forthcoming measurements of this mixing will play crucial role in
understanding of the underlying physics \cite{Mezzetto:2010zi}. Corrections to other angles
produce next order effect (as $s_{13}^2$), although in some cases
they can be enhanced by additional numerical factors.


\section{Acknowledgments}

The authors are grateful to M. Frigerio ans S. Khalil
for useful discussions.






\end{document}